\documentclass[aps,prd,twocolumn,superscriptaddress,longbibliography,amsmath,amssymb,amsfonts,citeautoscript]{revtex4-1}
\usepackage{amsmath}
\usepackage{amssymb}
\usepackage{bm}
\usepackage{bbm}
\usepackage{color}
\usepackage{epstopdf}
\usepackage{graphicx}
\usepackage[urlcolor=blue,colorlinks=true,citecolor=blue,linkcolor=blue,pdfstartview={FitH},bookmarks=false]{hyperref}
\usepackage{todonotes}
\usepackage{ulem}

\begin{document}

\title{Transient effects in double quantum dot sandwiched  laterally
       between \\superconducting and metallic leads}

\author{R. Taranko}
\affiliation{Institute of Physics, M. Curie-Sk\l{}odowska University,
             20-031 Lublin, Poland}

\author{K. Wrze\'sniewski}
\affiliation{Institute of Spintronics and Quantum Information, Faculty of Physics, Adam Mickiewicz University,
             61-614 Pozna\'n, Poland}

\author{B. Baran}
\affiliation{Institute of Physics, M. Curie-Sk\l{}odowska University,
             20-031 Lublin, Poland}

\author{I. Weymann}
\affiliation{Institute of Spintronics and Quantum Information, Faculty of Physics, Adam Mickiewicz University,
             61-614 Pozna\'n, Poland}

\author{T. Doma\'nski}
\affiliation{Institute of Physics, M. Curie-Sk\l{}odowska University,
             20-031 Lublin, Poland}

\date{\today}

\begin{abstract}
We study the transient phenomena appearing in a subgap region of the double quantum dot
coupled in series between the superconducting and normal metallic leads, focusing 
on the development of the superconducting proximity effect. For the uncorrelated 
nanostructure we derive explicit expressions of the time-dependent occupancies in 
both quantum dots, charge currents, and electron pairing induced on individual 
dots and between them. We show that the initial configurations substantially affect 
the dynamical processes, in which the in-gap bound states emerge upon coupling the 
double quantum dot to superconducting reservoir. In particular, the superconducting
proximity effect would be temporarily blocked whenever the quantum dots are initially 
singly occupied. Such {\it triplet}/{\it Andreev blockade} has been recently reported 
experimentally for double quantum dots embedded in the Josephson
[D. Bouman {\it et al.}, 
\href{https://link.aps.org/doi/10.1103/PhysRevB.102.220505}{Phys. Rev. B {\bf 102}, 220505 (2020)}]
and Andreev 
[P. Zhang {\it et al.}, \href{https://arxiv.org/abs/2102.03283}{arXiv:2102.03283 (2021)}]
junctions.
We also address the role of correlation effects within the lowest-order decoupling 
scheme and by the time-dependent numerical renormalization group calculations. 
Competition of the repulsive Coulomb interactions with the superconducting proximity 
effect leads to renormalization of the in-gap quasiparticles, speeding up the 
quantum oscillations and narrowing a region of transient phenomena, whereas 
the dynamical Andreev blockade is well pronounced in the weak inter-dot 
coupling limit. We propose feasible  methods for detecting the characteristic 
time-scales that could be observable by the Andreev spectroscopy.
\end{abstract}

\maketitle

\section{Introduction}
\label{sec.intro}

The transport of charge \cite{DeFranceschi-2010} and energy \cite{Zaikin-2012} through
heterostructures, where nanoscopic objects are attached to superconductor(s), is 
nowadays of great interest not only from the point of view of basic science
but, most importantly, due to promising future applications.
For instance, the quantum dots confined into Y-shape junction between 
two conducting and one superconducting electrode can be a source of spatially
entangled electrons from the dissociated Cooper pairs \cite{Hofstetter.2009}.
Another intensively studied field encompasses
semiconducting nanowires and/or magnetic nano-chains
hybridized with bulk superconductors, where the emerging topological phase
hosts Majorana quasiparticles, which are ideal candidates for stable qubits and
could enable quantum computations owing to their non-Abelian character \cite{Aasen-2016}.
These and many similar phenomena stem from the presence of bound states
that are induced at quantum dots/impurities \cite{balatsky.vekhter.06}, dimers \cite{Franke-2018},
nanowires \cite{Aguado.2017,Lutchyn.2018}, and magnetic nanoislands \cite{Menard.2017}
proximitized to bulk superconductors.

Since double quantum dot (DQD) configurations provide a versatile platform for 
the implementation of quantum information processing \cite{Kouwenhoven-2002,Nowack-2007},
such systems have also been considered in hybrid setups involving superconducting elements.
Experimentally, their bound states have been probed by the tunneling spectroscopy, using InAs \cite{Sherman.2017,Grove_Rasmussen.2018,Estrada_Saldana.2018,Estrada_Saldana.2020,Paaske-2020,Frolov-2021},
InSb \cite{Su.2017}, Ge/Si \cite{Zarassi.2017} quantum dots or carbon nanotubes \cite{Cleuziou.2006,Pillet.2013} 
contacted with superconducting lead(s), as well as by the scanning tunneling microscopy (STM) applied to the magnetic 
dimers deposited on superconducting substrates \cite{Ruby.2018,Franke-2018,Choi.2018,Kezilebieke.2019}.
The single V, Cr, Mn, Fe, and Co atoms deposited on aluminum have revealed that Cr and Mn atoms have 
contributions from different orbitals to subgap quasiparticles, whereas the other elements merely 
consist of one pair of the in-gap bound states \cite{Kustler-2021}. 
The properties of superconductor proximitized double quantum dots (dimers)
have been studied theoretically by a number groups \cite{Choi-2000,Zhu-2002,Tanaka.2010,Zitko.2010,Konig.2010,Rodero-11,Droste.2012,Grifoni.2013,Brunetti-2013,Yao2014Dec,Sothmann-2014,Trocha2015Jun,Meng.2015,Zitko-2015,Su.2017,Wrzesniewski2017Nov,Glodzik.2017,Frolov-2018,Scherubl_2019,Wojcik2019Jan,Zonda.2019,Wang-2019,Leijnse.2019}.
So far, however, hybrid DQD systems have been investigated mainly
under the stationary conditions \cite{balatsky.vekhter.06,Rodero-11},
while their transient behavior remains to a large extent unexplored.

\begin{figure}[t!]
\includegraphics[width=0.9\linewidth]{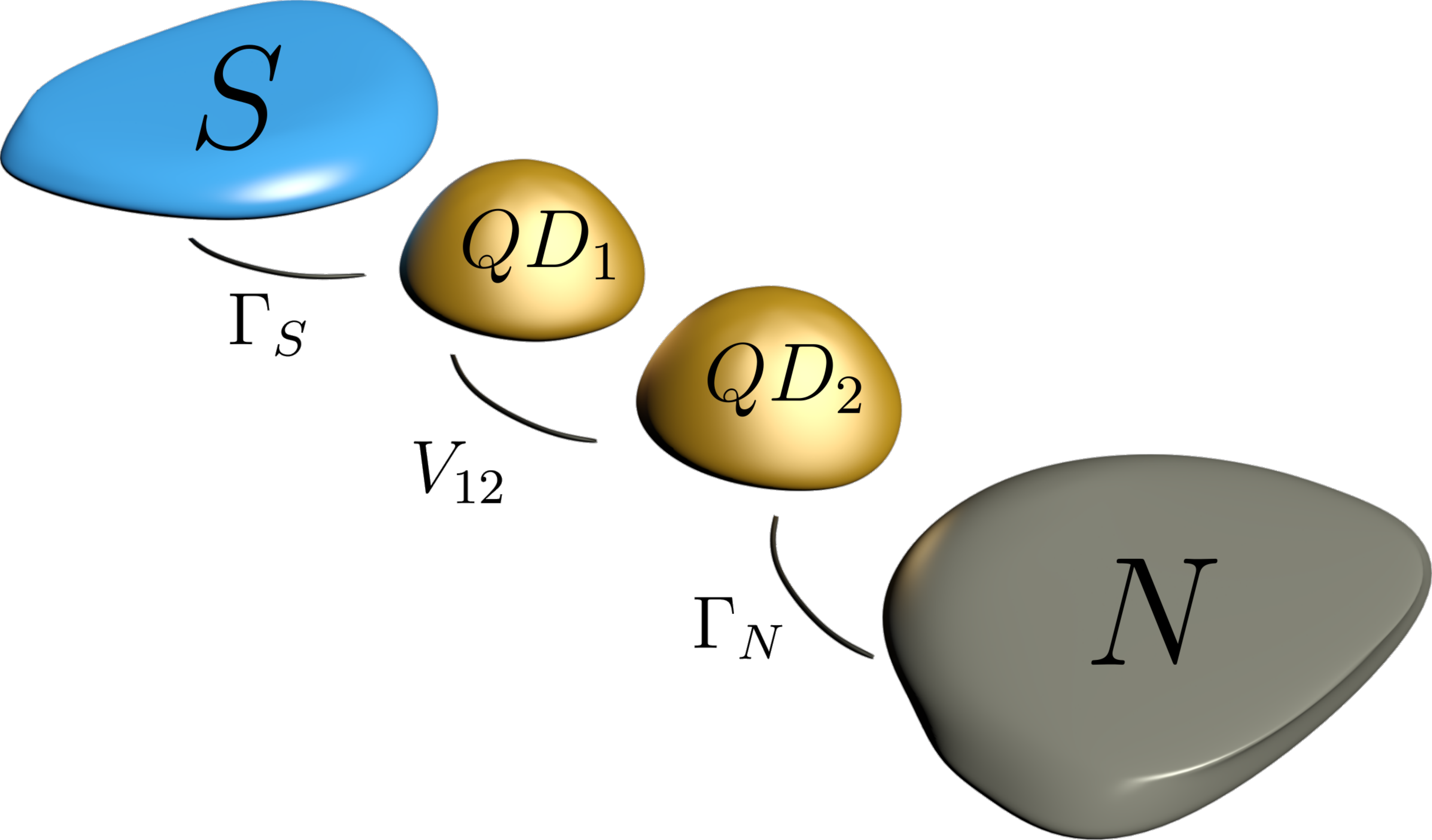}
\caption{Schematic view of the quantum dots (QD$_{1,2}$) embedded in series between the normal (N) and superconducting (S) leads with the couplings $\Gamma_{N}$ and $\Gamma_{S}$, respectively.}
\label{fig.scheme}
\end{figure}

In this paper we extend these studies by analyzing the dynamical phenomena after an abrupt attachment of DQD 
to the normal (N) and superconducting (S) electrodes (Fig.~\ref{fig.scheme}). We examine the development 
of the electron pairings on individual quantum dots as well as between them and analyze a gradual buildup 
of the subgap bound states. Our analytical expressions (obtained for uncorrelated setup) and numerical 
results (in the presence of Coulomb interactions) show that the initial configurations substantially affect 
the dynamical superconducting proximity effect. In particular, we reveal that the leakage of Cooper 
pairs onto both quantum dots would be blocked whenever the dots are initially singly occupied by the 
same spin electrons. This `{\it triplet}/{\it Andreev blockade}' has been recently observed 
experimentally under the stationary conditions, using DQD in the Josephson (S-DQD-S) 
\cite{Paaske-2020} and Andreev (N-DQD-S) junctions \cite{Frolov-2021}.
To get a deeper insight into the dynamical behavior in the considered N-DQD-S setup,
we analyze in detail various time-dependent quantities
taking into account several initial configurations.
To determine all relevant time-scales and examine the role of initial conditions,
we first derive the analytical results, neglecting the two-body interactions.
We then take into account the effects of the Coulomb repulsion using two different techniques.
To capture correlation effects, we make use of the mean-field approximation
to the Coulomb interaction, however, in further step we also employ
the time-dependent numerical renormalization group (tNRG) method
\cite{Wilson1975,Anders2005,Bulla2008},
which allows for obtaining very accurate predictions for the transient behavior
of an unbiased junction.
We demonstrate that the relevant time-scales are revealed in the transient currents
and could show up in other quench protocols, e.g.\ upon varying the quantum dot levels.

We believe that our study provides a valuable insight
into the dynamical superconducting proximity effect 
and the evolution of in-gap quasiparticles towards their stationary state values
in the case of double quantum dots. 
Our findings could be tested using the state of the art experimental techniques,
in particular, the subgap (Andreev) spectroscopy, and we hope
that this work will foster further efforts in studying dynamics of hybrid quantum dot structures.
Finally, we would like to notice that our analytical formalism can be extended to other quantum quench protocols,
for example, due to abrupt change of the quantum dots energy levels or periodic driving.
Moreover, it is also important to note that the presented analysis focuses on relatively weak coupling to the normal contact
and as such it does not encompass the subgap Kondo physics \cite{Tanaka.2010,Zitko-2015}.
This transport regime is definitely interesting and would require further analysis,
however, it goes beyond the scope of the present work.

The paper is organized as follows. In Sec.~\ref{sec.model_and_method} we introduce the 
microscopic model and describe the formalism for determination of time-dependent quantities. 
Section~\ref{finite_GammaN} presents the dynamics of the uncorrelated N-DQD-S setup, whereas 
Sec.~\ref{sec:Coulomb} is devoted to the studies of the role of the Coulomb interaction. In Sec.~\ref{sec:summary}
we summarize the main results and give a brief outlook. The technical details concerning 
the equations of motion of uncorrelated setup are presented in Appendix \ref{app:A}. 
In Appendix \ref{app:currents} we provide the expressions for the charge currents and 
Appendix~\ref{sec.dynamics} presents the analytical results for the uncorrelated DQD-S case.

\section{Formulation of the problem}
\label{sec.model_and_method}

\subsection{Microscopic model}
\label{sec.model}

The system under consideration (Fig.~\ref{fig.scheme}) consists of two quantum dots (QD$_{1,2}$)
placed in linear configuration between the superconducting (S) and normal (N) leads. The Hamiltonian of this
setup can be expressed as
\begin{equation}
\hat{H}=\hat{H}_{N}+\hat{H}_{S}+\hat{H}_{hybr}+\sum_{j=1,2} \hat{H}_{{QD}_{j}},
\label{Hamil_N_DQD_S}
\end{equation}
where $\hat{H}_{N}=\sum _{ {\bf{k}}}\varepsilon _{N {\textbf{k}}\sigma}\hat{c} _{N {\textbf{k}}\sigma}^{\dagger}\hat{c} _{N {\textbf{k}}\sigma}$
describes the normal lead electrons and the bulk superconductor is treated in the BCS-scenario
\begin{equation}
\hat{H}_{S}=\sum _{ {\textbf{q}}\sigma}\varepsilon _{S {\textbf{q}}}\hat{c} _{S {\textbf{q}}\sigma}^{\dagger}\hat{c} _{S {\textbf{q}}\sigma}
-\sum _{ {\textbf{q}}}\left( \Delta \hat{c}_{S {\textbf{q}}\uparrow}^{\dagger}\hat{c}_{S- {\textbf{q}}\downarrow}^{\dagger}+ \mbox{\rm h.c.} \right) .
\label{eq: 1}
\end{equation}
As usually,  $\hat{c}_{\beta {\textbf{k}}({\textbf{q}})\sigma}$ denote the second quantization operators
of the normal ($\beta$=N) and superconducting ($\beta$=S) lead electrons, respectively.
They are characterized by momenta $ {\textbf{k}} ( {\textbf{q}})$, energies $\varepsilon_{N {\textbf{k}} (S {\textbf{q}})}$ and
spin $\sigma=\uparrow,\downarrow$. We assume the pairing potential $\Delta$ of
superconducting lead to be real and restrict our considerations to the electronic states
inside this pairing gap window.

The external leads are interconnected via the quantum dots
$\hat{H}_{{QD}_{j}}=\sum_{\sigma}\varepsilon_{j\sigma}\hat{c}_{j\sigma}^{\dagger}\hat{c}_{j\sigma}$
whose energies are denoted by $\varepsilon_{j\sigma}$.
In our considerations we assume that the level spacing in the dots
is much larger than other energy scales, such that only a single orbital level
in each quantum dot is relevant for transport.
The constituents of the considered setup are hybridized through
\begin{eqnarray}
\hat{H}_{hybr}&=& \sum_{\sigma} \left(  V_{12}\hat{c}_{1\sigma}^{\dagger}\hat{c}_{2\sigma}
+ \sum _{ {\textbf{q}}}V _{S {\textbf{q}}}\hat{c} _{S {\textbf{q}}\sigma}^{\dagger}\hat{c}_{1\sigma}
\right.
\nonumber \\ &+& \left.
\sum _{ {\textbf{k}}}V _{N {\textbf{k}}}\hat{c} _{N {\textbf{k}}\sigma}^{\dagger}\hat{c}_{2\sigma}
+ \mbox{\rm h.c.} \right) ,
\label{eq: 2}
\end{eqnarray}
where $V_{12}$ denotes the inter-dot coupling, whereas $V_{S {\textbf{q}}(N {\textbf{k}})}$ describes
the coupling of QD$_{1(2)}$ to the external S(N) electrode. For convenience, we introduce the auxiliary
couplings $\Gamma_{\beta}=2\pi\sum_{ {\textbf{k}}} |V_{\beta {\textbf{k}}}|^{2} \delta(\omega - \varepsilon_{\beta {\textbf{k}}})$,
assuming them to be constant. Such constraint is realistic in the subgap region, $|\omega| < \Delta$, that is of our interest here.

For analytical determination of the time-dependent quantities we shall treat the pairing gap
$\Delta$ as the largest energy scale in this problem. Formally, we
thus focus on the superconducting atomic limit $\Delta\rightarrow\infty$. To simplify our notation
we set $\hbar=e=k_{B}=\Gamma_{S}=1$ when energies, currents and time are expressed in the units of $\Gamma_{S}$, $e\Gamma_{S}/\hbar$, and $\hbar/\Gamma_{S}$, respectively. In realistic situations $\Gamma_{S}\sim 200$ $\mu$eV, therefore the typical time-unit would be 3.3 psec and the current-unit $\sim 48$ nA.

\subsection{Transient evolution}
\label{sec.quench}

For $t < 0$ we assume all parts of the considered system to be disconnected.
The evolution of  the charge occupancies of quantum dots, $n_{j\sigma}(t)$, the transient currents flowing from the leads, $j_{N(S)\sigma}(t)$,  and the pairing correlation functions,  $\langle\hat{c}_{j\downarrow}(t)\hat{c}_{j\uparrow}(t)\rangle$ and $\langle\hat{c}_{1\downarrow}(t)\hat{c}_{2\uparrow}(t)\rangle$, driven by an abrupt hybridization (\ref{eq: 2}) at $t=0$ 
will bring the information about the superconducting proximity effect,
giving rise to the emergence of subgap quasiparticles.

The expectation value $\langle \hat{O}\rangle$ of any observable $\hat{O}$ can be determined by solving the Heisenberg 
equation of motion $i \frac{d}{dt}\hat{O}=\big[ \hat{O},\hat{H}\big]$. For this purpose it convenient to apply 
the Laplace transform
\begin{equation}
\hat{O}(s)=\int^{\infty}_{0} dt e^{-st}\hat{O}(t)
\label{Laplace_trabsform}
\end{equation}
to incorporate the initial ($t=0$) conditions \cite{Taranko-2018,Taranko-2019}. For example,
the time-dependent occupancy of the $j$-th QD would be formally given by
\begin{equation}
n_{j\sigma}(t)=\langle \mathcal{L}^{-1}\{\hat{c}_{j\sigma}^{\dagger}(s)\}(t)\cdot\mathcal{L}^{-1}\{\hat{c}_{j\sigma}(s)\}(t)  \rangle,
\label{QD_occup_def}
\end{equation}
where $\mathcal{L}^{-1}\{\hat{c}_{j\sigma}^{\dagger}(s)\}(t)$ stands for the  inverse Laplace transform of $\hat{c}_{j\sigma}^{\dagger}(s)$ and $\langle \cdots \rangle$ denotes the statistical averaging. 

When neglecting the Coulomb interactions on both quantum dots,
one can derive the explicit expressions for $\hat{c}_{j\sigma}^{(\dagger)}(s)$ and analytically determine the time-dependent expectation values of various observables (the influence of the correlation effects will be examined in Sec.\ \ref{sec:Coulomb}).
Let us now discuss the Laplace transforms of $\hat{c}_{j\sigma}(s)$, as they are crucial for the physical quantities of interest. Appendix \ref{app:A} presents the Laplace-transformed Heisenberg equations (\ref{A1}-\ref{A8}) for arbitrary value of the pairing gap $\Delta$. In the superconducting atomic limit ($\Delta \rightarrow \infty$) these equations simplify to
\begin{eqnarray}
(s\!+\!i\varepsilon_{1\uparrow})\hat{c}_{1\uparrow}(s)=-i\frac{\Gamma_{S}}{2}\hat{c}_{1\downarrow}^{\dagger}(s)-iV_{12}\hat{c}_{2\uparrow}(s)\!+\!\hat{a}_{1},
\label{eq:3}
\\
(s\!-\!i\varepsilon_{1\downarrow})\hat{c}_{1\downarrow}^{\dagger}(s)=-i\frac{\Gamma_{S}}{2}\hat{c}_{1\uparrow}(s)+iV_{12}\hat{c}_{2\downarrow}^{\dagger}(s)\!+\!\hat{a}_{2},
\label{eq:4}
\\
(s+i\varepsilon_{2\uparrow}+\frac{\Gamma_{N}}{2})\hat{c}_{2\uparrow}(s)=-iV_{12}\hat{c}_{1\uparrow}(s)+\hat{a}_{3},
\label{eq:5}
\\
(s-i\varepsilon_{2\downarrow}+\frac{\Gamma_{N}}{2})\hat{c}_{2\downarrow}^{\dagger}(s)=iV_{12}\hat{c}_{1\downarrow}^{\dagger}(s)+\hat{a}_{4}
\label{eq:6}
\end{eqnarray}
with $\hat{a}_{j}$ defined in (\ref{A13}-\ref{A19}). Here, we have used
\begin{eqnarray}
\sum_{\bf k}  \frac{\left| V_{N{\bf k}}\right|^{2}}{s\pm i \varepsilon_{N{\bf k}}}
= \frac{\Gamma_{N}}{2\pi} \int_{-D}^{D} \frac{d\omega}{s\pm i \omega} 
= \frac{\Gamma_{N}}{\pi} \arctan{ \left( \frac{D}{|s|} \right) } , 
\nonumber
\end{eqnarray}
which in the wide-bandwidth limit ($D \rightarrow \infty$)  implies $\sum _{ {\textbf{k}}}V _{N {\textbf{k}}}^{2}/(s \pm i\varepsilon _{N {\textbf{k}}}) \approx \Gamma_{N}/2$. In a similar way one finds
\begin{eqnarray}
\sum_{\bf q}  \left| V_{S{\bf q}}\right|^{2}\frac{s\pm i \varepsilon_{S{\bf q}}}
{s^{2} + \varepsilon_{S{\bf q}}^{2}+\Delta^{2}} 
\approx \frac{\Gamma_{S}}{2} \frac{s}{\sqrt{s^{2} +\Delta^{2}}} ,
\nonumber \\
\sum_{\bf q}  \left| V_{S{\bf q}}\right|^{2}\frac{\Delta}
{s^{2} + \varepsilon_{S{\bf q}}^{2}+\Delta^{2}}
\approx
\frac{\Gamma_{S}}{2} \frac{\Delta}{\sqrt{s^{2} +\Delta^{2}}} .
\nonumber
\end{eqnarray}
Thus, in the superconducting atomic limit we  have 
 $\lim_{\Delta\rightarrow\infty}\sum _{ {\textbf{q}}}|V _{S {\textbf{q}}}|^{2}(s\pm i\varepsilon _{S {\textbf{q}}})/(s^{2}+\varepsilon _{S {\textbf{q}}}^{2}+\Delta^{2})\approx 0$  and $\lim_{\Delta\rightarrow\infty}\sum _{ {\textbf{q}}}|V _{S {\textbf{q}}}|^{2}\Delta/(s^{2}+\varepsilon _{S {\textbf{q}}}^{2}+\Delta^{2})\approx \Gamma_{S}/2$, respectively.

For the specific case of $\varepsilon_{j\sigma}=0$, the Laplace transforms of QD operators
can be expressed as
%
\onecolumngrid
\begin{eqnarray}
\hat{c}_{1\uparrow/\downarrow}(s) &=&  \hat{c}_{1\uparrow/\downarrow}(0) \frac{u(s)(s+\Gamma_{N}/2)}{W(s)}
-\hat{c}_{2\uparrow/\downarrow}(0)\frac{iV_{12} u(s)}{W(s)}
\mp \hat{c}_{1\downarrow/\uparrow}^{\dagger}(0)\frac{\Gamma_{S}(s+\Gamma_{N}/2)^{2}}{2W(s)}
\pm \hat{c}_{2\downarrow/\uparrow}^{\dagger}(0) \frac{\Gamma_{S}(s+\Gamma_{N}/2)V_{12}}{2W(s)}
\nonumber \\
&-& \sum _{ {\textbf{k}}} \left[ 
\hat{c}_{N {\textbf{k}}\uparrow/\downarrow}(0) \frac{V _{N {\textbf{k}}}}{s+i\varepsilon _{N {\textbf{k}}}} \frac{V_{12} u(s)}{W(s)} \mp
\hat{c}_{N {\textbf{k}}\downarrow/\uparrow}^{\dagger}(0)\frac{V _{N {\textbf{k}}}}{s-i\varepsilon _{N {\textbf{k}}}}\frac{i\Gamma_{S}V_{12}(s+\Gamma_{N}/2)}{2W(s)}
 \right]
\nonumber \\
&- &
\frac{i\Gamma_{S}}{2W(s)}\sum _{ {\textbf{q}}} \left[ \frac{V _{S {\textbf{q}}}\Delta u(s)}{s^{2}+\varepsilon _{S {\textbf{q}}}^{2}+\Delta^{2}}\hat{c}_{S- {\textbf{q}}\uparrow/\downarrow}(0)\pm i \frac{V _{S {\textbf{q}}} u(s)(s+i\varepsilon _{S {\textbf{q}}})}{s^{2}+\varepsilon _{S {\textbf{q}}}^{2}+\Delta^{2}}\hat{c}_{S {\textbf{q}}\downarrow/\uparrow}^{\dagger}(0)\right]
\nonumber \\
&+& \frac{u(s)(s+\Gamma_{N}/2)}{W(s)} \left[\mp\sum _{ {\textbf{q}}}\frac{V _{S {\textbf{q}}}\Delta}{s^{2}+\varepsilon _{S {\textbf{q}}}^{2}+\Delta^{2}}\hat{c}_{S- {\textbf{q}}\downarrow/\uparrow}^{\dagger}(0) -i\sum _{ {\textbf{q}}}\frac{V _{S {\textbf{q}}}(s-i\varepsilon _{S {\textbf{q}}})}{s^{2}+\varepsilon _{S {\textbf{q}}}^{2}+\Delta^{2}}\hat{c}_{S {\textbf{q}}\uparrow/\downarrow}(0)\right] ,
\label{eq:7}
\\
 \hat{c}_{2\uparrow/\downarrow}(s) &=& - \hat{c}_{1\uparrow/\downarrow}(0)\frac{iV_{12} u(s)}{W(s)}
+\hat{c}_{2\uparrow/\downarrow}(0)\frac{1-V^{2}_{12} u(s)/W(s)}{s+\Gamma_{N}/2}
\mp \hat{c}_{1\downarrow/\uparrow}^{\dagger}(0) \frac{V_{12} \Gamma_{S}(s+\Gamma_{N}/2)}{2W(s)}
\mp\hat{c}_{2\downarrow/\uparrow}^{\dagger}(0) \frac{ i V_{12}^{2}\Gamma_{S}}{2 W(s)}
\nonumber \\
&+ & 
\sum _{ {\textbf{k}}}\left[ \hat{c}_{N {\textbf{k}}\uparrow/\downarrow}(0)\frac{V _{N {\textbf{k}}}}{s+i\varepsilon _{N {\textbf{k}}}}\frac{V_{12}^{2} u(s)/W(s)-1}{s+\Gamma_{N}/2} \pm \hat{c}_{N {\textbf{k}}\downarrow/\uparrow}^{\dagger}(0)\frac{V _{N {\textbf{k}}}}{s-i\varepsilon _{N {\textbf{k}}}} \frac{ V_{12}^{2} \Gamma_{S}}{2 W(s)} \right]
\nonumber \\
&-& \frac{V_{12}\Gamma_{S}(s+\Gamma_{N}/2)}{2W(s)} \left[ \sum _{ {\textbf{q}}} \frac{V _{S {\textbf{q}}}\Delta}{s^{2}+\varepsilon _{S {\textbf{q}}}^{2}+\Delta^{2}}\hat{c}_{S- {\textbf{q}}\uparrow/\downarrow}(0)\pm i\sum _{ {\textbf{q}}} \frac{V _{S {\textbf{q}}}(s+i\varepsilon _{S {\textbf{q}}})}{s^{2}+\varepsilon _{S {\textbf{q}}}^{2}+\Delta^{2}}\hat{c}_{S {\textbf{q}}\downarrow/\uparrow}^{\dagger}(0)\right]
\nonumber \\
&-& \frac{i V_{12} u(s)}{W(s)} \left[\mp\sum _{ {\textbf{q}}}\frac{V _{S {\textbf{q}}}\Delta}{s^{2}+\varepsilon _{S {\textbf{q}}}^{2}+\Delta^{2}}\hat{c}_{S- {\textbf{q}}\downarrow/\uparrow}^{\dagger}(0) -i\sum _{ {\textbf{q}}}\frac{V _{S {\textbf{q}}}(s-i\varepsilon _{S {\textbf{q}}})}{s^{2}+\varepsilon _{S {\textbf{q}}}^{2}+\Delta^{2}}\hat{c}_{S {\textbf{q}}\uparrow/\downarrow}(0)\right] ,
\label{eq:8}
\end{eqnarray}
\twocolumngrid
\noindent
where
\begin{eqnarray}
u(s) &=& s \left( s+\Gamma_{N}/2 \right) +V_{12}^{2}, \label{omega_s}\label{eq:9} \\
W(s) &=& u^{2}(s)+\left( \frac{\Gamma_{S}}{2} \right)^{2}\left( s+\frac{\Gamma_{N}}{2}\right)^{2}.
\label{eq:10}
\end{eqnarray}
In Eqs.~(\ref{eq:7},\ref{eq:8}) there appears the pairing gap $\Delta$, originating from the auxiliary operators $\hat{a}_{j}$. We impose the superconducting atomic limit values  later on, when computing the statistically averaged observables \cite{Taranko-2018}.

The 4-th order polynomial (\ref{eq:10}) can be rewritten as, $W(s)=s^{4}+b_{3}s^{3}+b_{2}s^{2}+b_{1}s+b_{0}$, with the real coefficients,
$b_{0}=V_{12}^{4}+\Gamma_{S}^{2}\Gamma_{N}^{2}/16$, $b_{1}=\Gamma_{N}V_{12}^{2}+\Gamma_{N}\Gamma_{S}^{2}/4$, $b_{2}=2V_{12}^{2}+(\Gamma_{S}^{2}+\Gamma_{N}^{2})/4$ and $b_{3}=\Gamma_{N}$.
It can be recast into a product $W(s)=(s-s_{1})(s-s_{2})(s-s_{3})(s-s_{4})$, whose roots  obey $s_{2}=s_{1}^{*}$ and $s_{3}=s_{4}^{*}$. Their knowledge enables us to obtain the inverse Laplace transforms of $\hat{c}_{j\sigma}(s)$ operators,  expressing the time-dependent charge occupancies $n_{j\sigma}(t)$, pairing correlation functions, $\langle \hat{c}_{j-\sigma}(t) \hat{c}_{j\sigma}(t)\rangle$, $\langle \hat{c}_{1\downarrow}(t) \hat{c}_{2\uparrow}(t)\rangle$, and transient currents induced between various sectors of the N-DQD-S setup.

\section{Dynamics of uncorrelated setup}
\label{finite_GammaN}

In this section we analyze the time-dependent observables obtained analytically for N-DQD-S nanostructure
by the equation of motion procedure in the absence of the Coulomb repulsion.
We begin by discussing the electron occupancies of each QD derived by inserting the inverse
Laplace transforms [Eqs.~(\ref{eq:7})-(\ref{eq:8})] to Eq.\ (\ref{QD_occup_def}).
For $t < 0$, all parts of the setup are disconnected, therefore the occupancy $n_{j\sigma}(t>0)$
consists of the contributions from the initial expectation values of $n_{j\sigma}(0)$, $\langle \hat{c}^{\dagger} _{N {\textbf{k}}\sigma}(0)\hat{c} _{N {\textbf{k}}\sigma}(0)\rangle$, $\langle \hat{c}^{\dagger} _{S {\textbf{q}}\sigma}(0)\hat{c} _{S {\textbf{q}}\sigma}(0)\rangle$, and $\langle \hat{c} _{S {\textbf{q}}\sigma}^{(\dagger)}(0)\hat{c}_{S- {\textbf{q}}\bar{\sigma}}^{(\dagger)}(0)\rangle$, where $\bar{\sigma}$ is opposite spin to $\sigma$. In the superconducting atomic limit, for  $t>0$, we obtain 
\onecolumngrid
\begin{eqnarray}
n_{1\uparrow/\downarrow}(t)&=&n_{1\uparrow/\downarrow}(0) \left( \mathcal{L}^{-1} \left\{ \frac{ u(s)(s+\Gamma_{N}/2)}{W(s)}  \right\} (t) \right) ^{2} +n_{2\uparrow/\downarrow}(0)V_{12}^{2} \left( \mathcal{L}^{-1} \left\{ \frac{u(s)}{W(s)} \right\} (t) \right) ^{2} \nonumber \\
&+&(1-n_{1\downarrow/\uparrow}(0))\Gamma_{S}^{2}/4 \left( \mathcal{L}^{-1} \left\{ \frac{ (s+\Gamma_{N}/2)^{2}}{W(s)} \right\} (t) \right) ^{2} +(1-n_{2\downarrow/\uparrow}(0))V_{12}^{2}\Gamma_{S}^{2}/4 \left( \mathcal{L}^{-1} \left\{ \frac{ s+\Gamma_{N}/2}{W(s)} \right\} (t) \right) ^{2} \nonumber \\
&+& V_{12}^{2}\frac{\Gamma_{N}}{2\pi}\int_{-\infty}^{\infty}d\varepsilon f_{N}(\varepsilon)\mathcal{L}^{-1} \left\{ \frac{u(s)}{(s+i\varepsilon)W(s)} \right\}(t)\cdot  \mathcal{L}^{-1} \left\{ \frac{u(s)}{(s-i\varepsilon)W(s)}  \right\}(t) \nonumber \\
&+&\Gamma_{N}\Gamma_{S}^{2}V_{12}^{2}/8\pi\int_{-\infty}^{\infty}d\varepsilon(1-f_{N}(\varepsilon))\mathcal{L}^{-1} \left\{ \frac{s+\Gamma_{N}/2}{(s+i\varepsilon)W(s)}  \right\}(t)\cdot \mathcal{L}^{-1} \left\{ \frac{s+\Gamma_{N}/2}{(s-i\varepsilon)W(s)}  \right\}(t),
\label{eq:B1}
\end{eqnarray}
\begin{eqnarray}
n_{2\uparrow/\downarrow}(t)&=&n_{1\uparrow/\downarrow}(0)V_{12}^{2} \left( \mathcal{L}^{-1} \left\{ \frac{u(s)}{W(s)} \right\} (t) \right) ^{2} +n_{2\uparrow/\downarrow}(0) \left( \mathcal{L}^{-1} \left\{ \frac{1}{s+\Gamma_{N}/2}-\frac{u(s)V_{12}^{2}}{W(s)(s+\Gamma_{N}/2)}  \right\} (t) \right) ^{2} \nonumber \\
&+&(1-n_{1\downarrow/\uparrow}(0))V_{12}^{2}\Gamma_{S}^{2}/4 \left( \mathcal{L}^{-1} \left\{ \frac{ s+\Gamma_{N}/2}{W(s)} \right\} (t) \right) ^{2} +(1-n_{2\downarrow/\uparrow}(0))V_{12}^{4}\Gamma_{S}^{2}/4 \left( \mathcal{L}^{-1} \left\{ \frac{1}{W(s)} \right\} (t) \right) ^{2} \nonumber \\
&+&\frac{\Gamma_{N}}{2\pi}\int_{-\infty}^{\infty}d\varepsilon f_{N}(\varepsilon)\mathcal{L}^{-1} \left\{ (\frac{V_{12}^{2} u(s)}{W(s)}-1 )\frac{1}{(s-i\varepsilon)(s+\Gamma_{N}/2)}\right\}(t)\cdot  \mathcal{L}^{-1} \left\{ (\frac{V_{12}^{2} u(s)}{W(s)}-1)(\frac{1}{(s+i\varepsilon)(s+\Gamma_{N}/2)})  \right\}(t) \nonumber \\
&+&\Gamma_{N}\Gamma_{S}^{2}V_{12}^{4}/8\pi\int_{-\infty}^{\infty}d\varepsilon(1-f_{N}(\varepsilon))\mathcal{L}^{-1} \left\{ \frac{1}{(s+i\varepsilon)W(s)}  \right\}(t)\cdot \mathcal{L}^{-1} \left\{ \frac{1}{(s-i\varepsilon)W(s)}  \right\}(t),
\label{eq:B2}
\end{eqnarray}
\twocolumngrid
\noindent
where $f_{N}(\varepsilon)=\left[1+\mbox{\rm exp}\left(\varepsilon/k_{B}T\right)\right]^{-1}$.
The time-dependent occupancies depend on the initial DQD configurations $n_{j}(0)$ [through the first four terms appearing in Eqs.~(\ref{eq:B1},\ref{eq:B2})] and on the couplings to external leads (via the last two terms).
Let us notice that for the initial triplet configuration ($n_{j\uparrow}(0)=0$, $n_{j\downarrow}(0)=1$)
the evolution of $n_{j\sigma}(t)$ would be solely controlled by the coupling $\Gamma_{N}$ to metallic lead.

\begin{figure}[t]
\centerline{\includegraphics[width=1\linewidth]{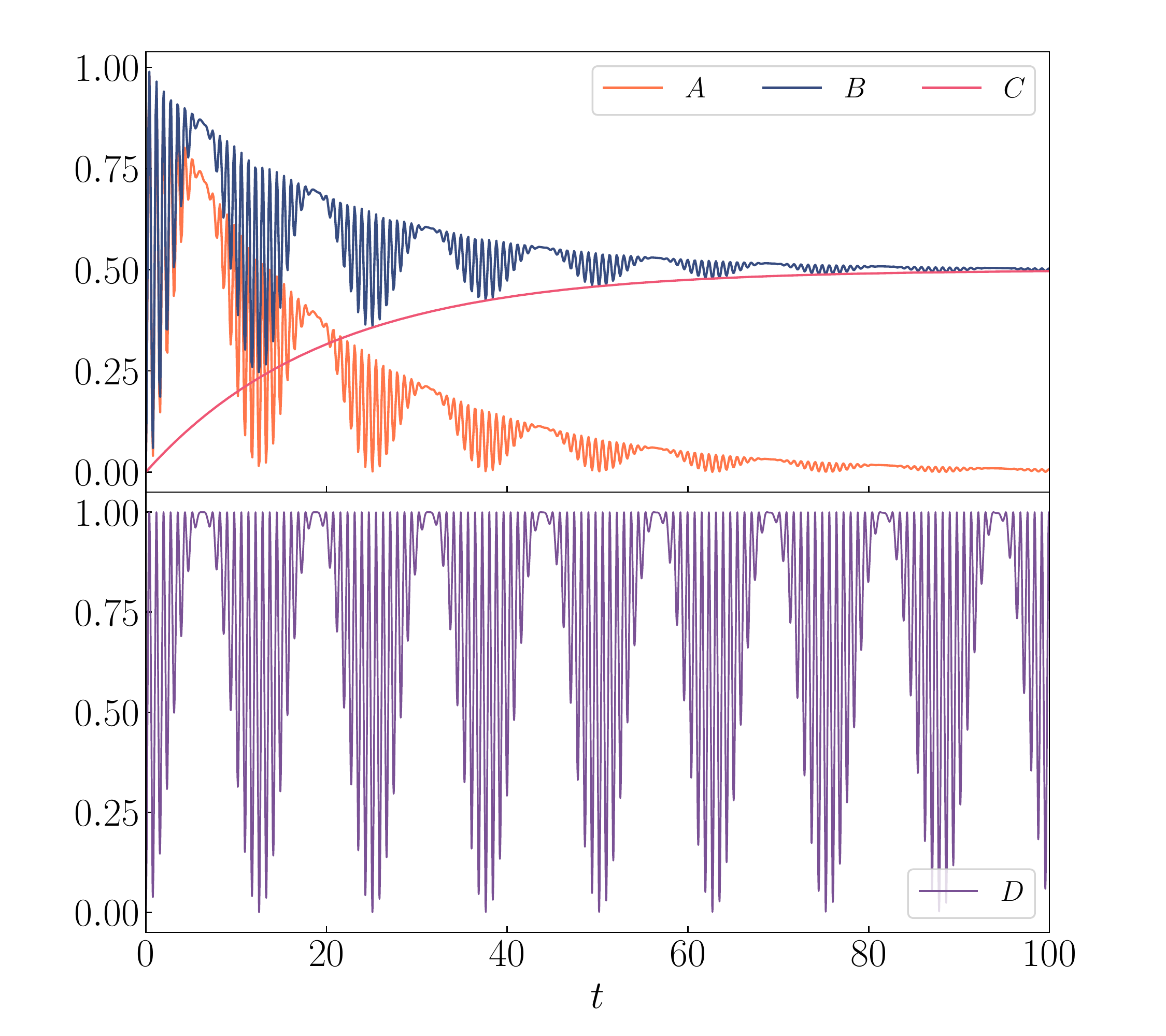}}
\caption{Comparison of $n_{2\uparrow}(t)$ obtained for $\Gamma_{N}=0.1\Gamma_{S}$ (curve B) with the case $\Gamma_{N}=0$ (curve D). The curve C shows the contribution to $n_{2\uparrow}(t)$ due to the coupling $\Gamma_{N}$ and the curve A refers to the contribution strictly dependent on the initial occupancies. Calculations have been done for $V_{12}=4\Gamma_{S}$, assuming the initial conditions (QD$_{1}$,QD$_{2}$)=($\uparrow,0$), $\Gamma_{S}=1.0$, $\Gamma_{N}=0.1$, $\varepsilon_{j\sigma}=0$.}
\label{fig6}
\end{figure}

\begin{figure}[t]
\centerline{\includegraphics[width=1\linewidth]{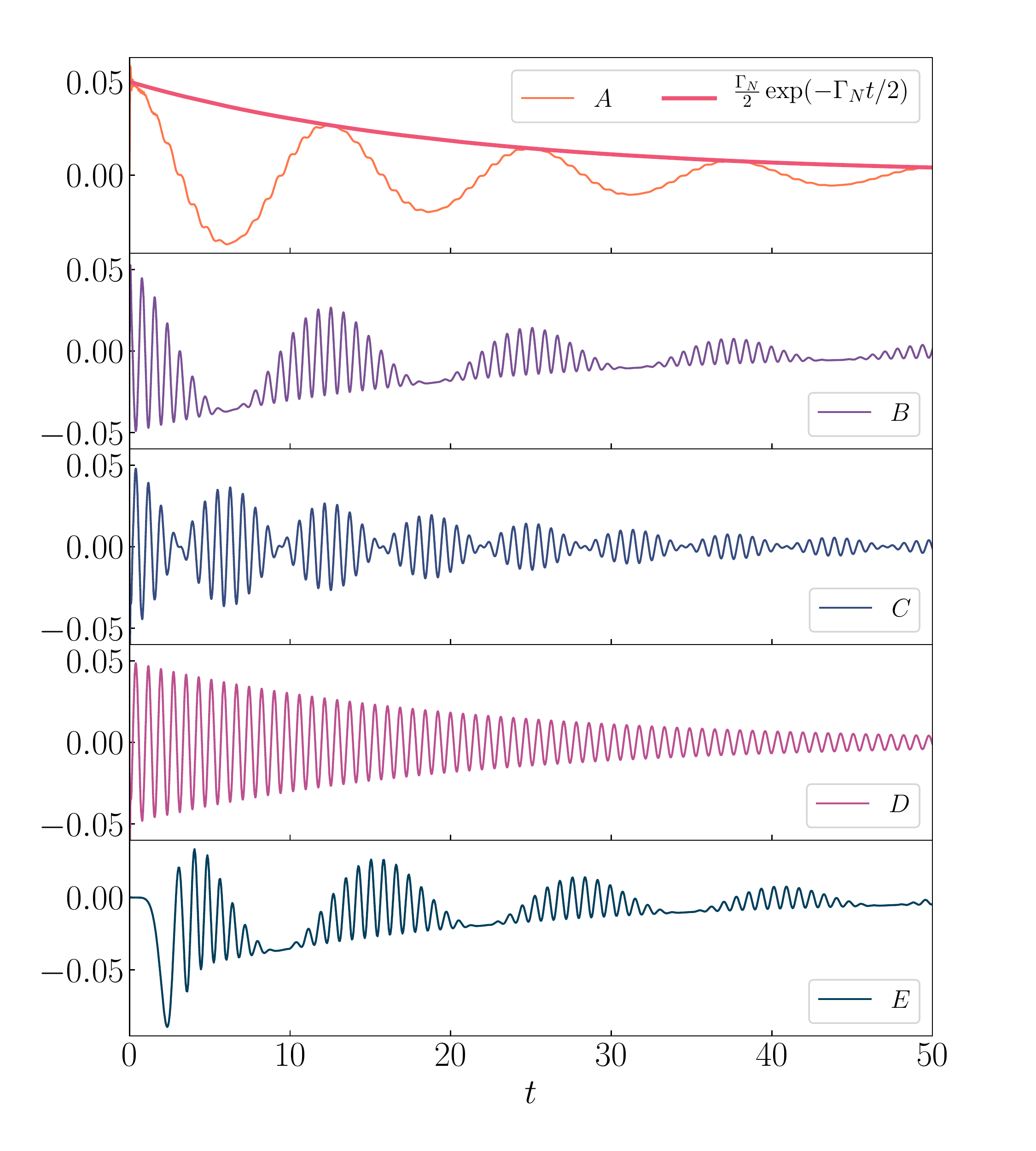}}
\caption{The time-dependent current $j_{N\uparrow}(t)$ obtained for several initial configurations A=($0,0$), B=($0,\uparrow$), C=($\uparrow\downarrow,0$), and D=($\uparrow$,$\downarrow$). The thick solid line shows the envelope function $\frac{\Gamma_{N}}{2}\exp{(-\Gamma_{N}t/2)}$, which refers to the damped currents obtained for all initial configurations.
	The bottom curve shows the current obtained for a finite interval of the switching time (see the text) of couplings between various constituents of the system,
	which can be compared with the curve B. Numerical results are obtained for $V_{12}=4\Gamma_{S}$,
	$\Gamma_{N}=0.1\Gamma_{S}$ and $\varepsilon_{j\sigma}=0$.}
\label{fig7}
\end{figure}

As an example, in Fig.~\ref{fig6} we show the time-dependent occupancy of the second dot $n_{2\uparrow}(t)$
obtained for a strong interdot coupling $V_{12}=4\Gamma_{S}$ in the unbiased heterostructure ($\mu_{N}=0=\mu_{S}$), assuming that initially only spin-$\uparrow$ electron occupies the first QD, (QD$_{1}$,QD$_{2}$)=($\uparrow,0$).
For comparison, in the bottom panel we display the results in the absence of the metallic lead. We also present the contributions described by the first four terms of the general formula Eq.~(\ref{eq:B2}), which are dependent on the initial occupancies.
We can notice the oscillating character of $n_{2\uparrow}(t)$ with a damping imposed by $\Gamma_{N}$. The stationary limit value $n_{2\uparrow}(t\rightarrow\infty) = 0.5$ is approached through a sequence of quantum oscillations whose amplitude is exponentially suppressed with an envelope function $\exp(-\Gamma_{N}t/2)$.
Such behavior is a consequence of the superposition of 
damped transient oscillations and another part which is independent of the initial occupancies [expressed by the last two terms in Eq.\ (\ref{eq:B2})]
arising from the direct coupling of QD$_{2}$ to the normal lead. In the case of unbiased junction, the latter part simplifies to $\frac{1}{2} \left( 1-\exp(\Gamma_{N}t/2)\right)$, as displayed by C curve in Fig.~\ref{fig6}.

We now consider the subgap (Andreev) current $j_{N\sigma}(t)$, flowing from the normal lead to QD$_{2}$
\begin{equation}
j_{N\sigma}(t) = 2 \mbox{\rm Im} \sum _{ {\textbf{k}}} V _{N {\textbf{k}}} \langle \hat{c}^{\dagger}_{2\sigma}(t) \hat{c} _{N {\textbf{k}}\sigma}(t) \rangle .
\label{current_N}
\end{equation}
In the wide bandwidth limit it can be expressed as \cite{Taranko-2018}
\begin{equation}
j_{N\sigma}(t) = 2 \mbox{\rm Im} \sum _{ {\textbf{k}}} V _{N {\textbf{k}}} e^{-i\varepsilon _{N {\textbf{k}}}t}\langle \hat{c}^{\dagger}_{2\sigma}(t)\hat{c} _{N {\textbf{k}}\sigma}(0) \rangle - \Gamma_{N} n_{2\sigma}(t).
\end{equation}
Using the Hermitian conjugate of the operator $\hat{c}_{2\sigma}(t)$ presented in Eq.~(\ref{eq:8}), we explicitly obtain
%
\begin{eqnarray}
j_{N\sigma}(t) &=& -\Gamma_{N}n_{2\sigma}(t) + \frac{\Gamma_{N}}{\pi}
\mbox{\rm Re} \left\{ \int^{\infty}_{-\infty}d\varepsilon f_{N}(\varepsilon) e^{-i\varepsilon t}
\right. \nonumber \\
&\times & \left[ \mathcal{L}^{-1}\left\{ \frac{1}{(s-i\varepsilon)(s+g_{n})} \right\}(t)
\right. \nonumber \\
&-& \left. \left.
\mathcal{L}^{-1} \left\{ \frac{u(s)V^{2}_{12}}{(s-i\varepsilon)(s+g_{n})W(s)}  \right\}(t) \right] \right\} ,
\label{j_N}
\end{eqnarray}
%
where the time-dependent occupation $n_{2\sigma}(t)$ is given by Eq.\ (\ref{eq:B2}).

Figure \ref{fig7} displays the Andreev current $j_{N\uparrow}(t)$ computed for representative initial configurations, namely:  
A=($0,0$), B=($0,\uparrow$), C=($\uparrow\downarrow,0$), and D=($\uparrow$,$\downarrow$). The quantum oscillations appearing 
in $j_{N\uparrow}(t)$ are identical with the time-dependent variation of the occupancies $n_{j\uparrow}(t)$ of the simpler 
DQD-S case [see Fig.~\ref{fig2} in Appendix \ref{sec.dynamics}] and the supercurrent $j_{S\sigma}(t)$ [Fig.~\ref{fig4}]. Here the main difference  refers to the relaxation processes, which impose a damping on such quantum oscillations. This effect can be described by the envelope function $\frac{\Gamma_{N}}{2}\exp(-\Gamma_{N}t/2)$ (see Fig.~\ref{fig7}).
Apart from this damping, all other features appearing in $n_{j\sigma}(t)$ and $j_{S\sigma}(t)$ (e.g.\ oscillations with the periods of $\pi/V_{12}$ and $4\pi/\Gamma_{S}$) are present in the time-dependent Andreev current $j_{N\sigma}(t)$, too.

Let us now comment on the large value of the transient current $j_{N\sigma}(0^{+})$  right after forming the N-DQD-S setup (Fig.~\ref{fig7}). Such rapid increase of the current from zero to $\frac{\Gamma_{N}}{2}$ is unphysical and in realistic experimental situations would not occur \cite{Komnik.2008}. We have checked numerically that this artifact is absent for the smooth in time coupling protocol, $V_{N {\textbf{k}}/S {\textbf{q}}}(t)=\frac{V_{N {\textbf{k}}/ {S\textbf{q}}}}{2}\left[ \sin\left( \pi \left( \frac{t}{t^{*}}  - \frac{1}{2} \right) \right) +1 \right]$ for $0<t\leq t^{*}$, and next imposing the constant value $V_{N {\textbf{k}}/S {\textbf{q}}}(t>t^{*})=V_{N {\textbf{k}}/S {\textbf{q}}}$. The bottom (E) panel of Fig.~\ref{fig7} presents  the transient current obtained for $t^{*}=5$. Indeed, the absolute value of  $|j_{N\uparrow}(t)|$ continuously increases from zero. Its variation in time in the region of $t\geq t^{*}$ is roughly the same as for the abrupt switching of coupling. Similar tendency holds  for other quantities as well.

\begin{figure*}
\centerline{\includegraphics[width=1\linewidth]{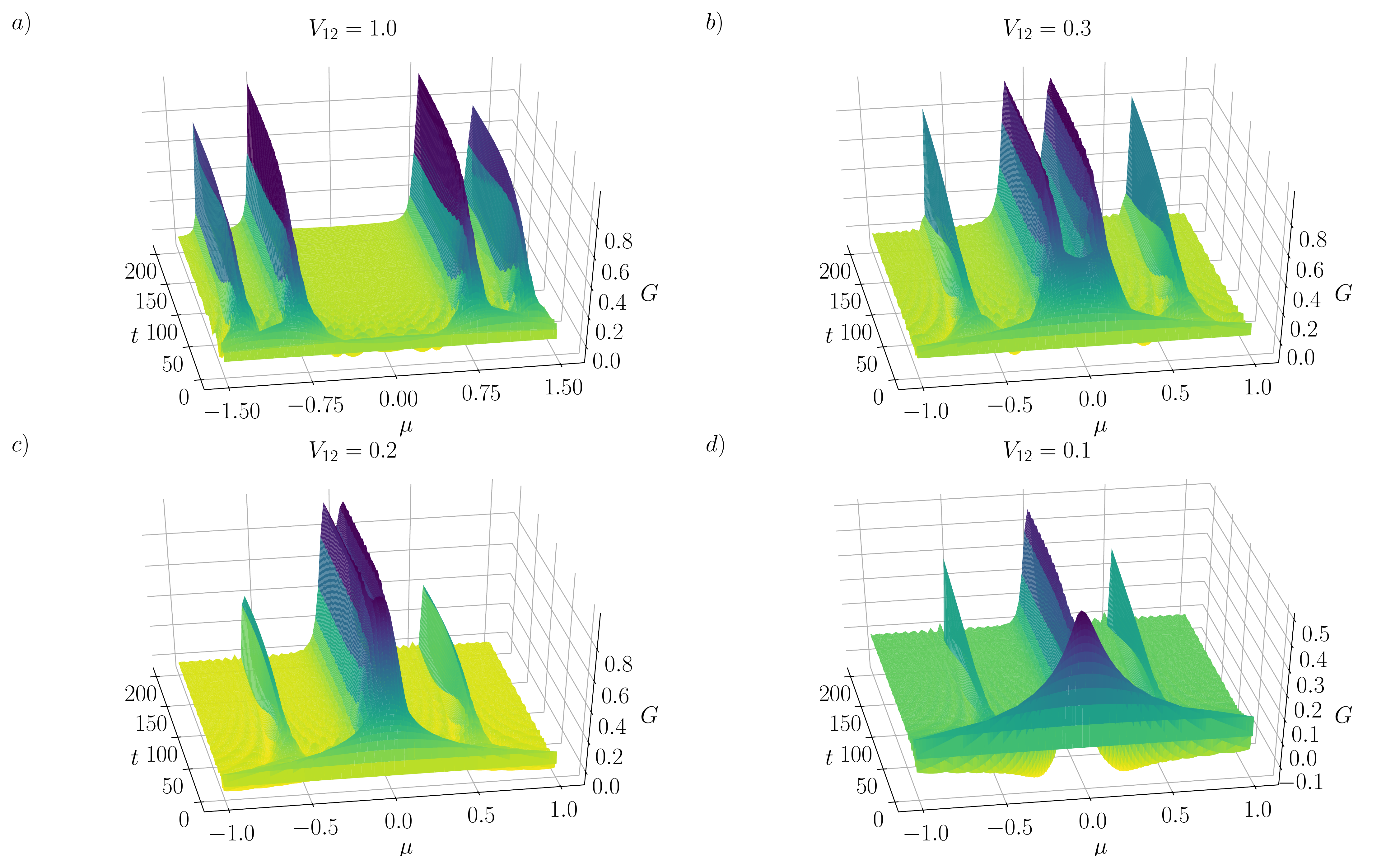}}
\caption{The differential conductance $G$ (in units of $2e^{2}/h$) as a function of time $t$ and the bias voltage $\mu$ obtained for $V_{12}=1$, $0.3$, $0.2$ and $0.1$ (A, B, C and D panels, respectively), assuming $\Gamma_{S}=1$, $\Gamma_{N}=0.1$ and $\varepsilon_{j\sigma}=0$.}
\label{fig8}
\end{figure*}

Using  the expression (\ref{j_N}) for $j_{N\sigma}(t)$ we define its time-dependent differential conductance $G_{\sigma}(\mu,t)=\frac{d}{d\mu}j_{N\sigma}(t)$ as a function of the bias voltage $\mu \equiv \mu_{N}- \mu_{S}$. At zero temperature it takes the following form (in units of $2e^{2}/h$)
%
\onecolumngrid
\begin{eqnarray}
&&G_{\sigma}(\mu,t)
= \Gamma_{N} \mbox{\rm Re} \left\{ \exp(-i\mu t) \left[ \mathcal{L}^{-1}\left\{ \frac{1}{(s-i\mu)(s+g_{n})}\right\} (t)
- V^{2}_{12}\mathcal{L}^{-1}\left\{ \frac{u(s)}{(s-i\mu)(s+g_{n})W(s)}\right\} (t) \right] \right. \label{eq49} \\
&&- \left. \frac{\Gamma^{2}_{N}}{2} \left| \mathcal{L}^{-1}\left\{ \frac{1}{(s+i\mu)(s+g_{n})}\right\} (t)
-V^{2}_{12}\mathcal{L}^{-1}\left\{  \frac{u(s)}{(s+i\mu)(s+g_{n})W(s)}\right\} (t) \right|^{2}
+\frac{\Gamma^{2}_{N}\Gamma^{2}_{S}V^{4}_{12}}{8} \left| \mathcal{L}^{-1}\left\{ \frac{1}{(s+i\mu)W(s)}\right\} \right|^2 \right\} .
\nonumber
\end{eqnarray}
\twocolumngrid
%
\noindent
Note, that for the specific case of $\varepsilon_{j\sigma}=0$, the differential conductance
is spin-independent, $G_{\sigma}(\mu,t)=G(\mu,t)$.  

The peaks appearing in the conductance $G(\mu,t)$ as a function of $\mu$ can be identified as the quasiparticle excitation energies between  eigenstates comprising the even and odd numbers of electrons. For the uncorrelated DQD these bound states occur at $E_{A}=\pm\frac{1}{2}(\sqrt{4V^{2}_{12}+\Gamma^{2}_{S}/4}\pm\frac{\Gamma_{S}}{2})$. We have calculated numerically the conductance  (\ref{eq49}) and observed the emergence of such bound states at $E_{A}$ upon approaching the steady limit $t\rightarrow\infty$. In our N-DQD-S heterostructure they acquire a broadening (i.e.\ finite life-time) due to scattering on a continuous spectrum of the normal lead. Electronic states from outside the pairing gap of superconducting lead could additionally broaden these bound states, supporting the relaxation mechanism \cite{LevyYeyati-2017}.

In Fig.~\ref{fig8} we plot the differential conductance  vs time and bias voltage for several interdot couplings, $V_{12}$, ranging from the large (panel a) to small (panel d) values.
For $V_{12}\geq \Gamma_{S}$, we observe the emergence of two broad structures at early times  around the quasiparticle energies $\pm V_{12}$. In a short time-period $\Delta t$ right after the quench (here $\Delta t \sim 10$), these features evolve into distinct peaks, separated by $\sim\Gamma_{S}/2$. By decreasing $V_{12}$, we observe that  the low energy excitations  are merged until certain  time after quench, whereas the higher energy excitations are well separated  from each other. With further decrease of $V_{12}$ the energy difference between the low energy excitations disappears. For very small $V_{12}=0.1\Gamma_{S}$ [see Fig.~\ref{fig8}(d)], the low energy excitations form a single broad peak at zero energy, coexisting with the side-attached excitations at energies $\pm\Gamma_{S}/2$ of low intensity. This structure emerges at late times ($\sim 60$) after the quench.

Finally, we examine the relationship between the excitation energies obtained from the differential conductance and  the time-dependent electron pairings induced on individual quantum dots, $\langle c_{j\downarrow}(t)  c_{j\uparrow}(t) \rangle$, and between them $\langle c_{1\downarrow}(t)  c_{2\uparrow}(t) \rangle$. Using the inverse Laplace transforms of the operators $c_{j\sigma}(s)$ we get 
%
\onecolumngrid
\begin{eqnarray}
\langle \hat{c}_{1\downarrow}(t)\hat{c}_{1\uparrow}(t)\rangle &=& i\Gamma_{S}/2 \left[ (n_{1\uparrow}(0)+n_{1\downarrow}(0)-1)\mathcal{L}^{-1} \left\{ \frac{ (s+\Gamma_{N}/2)^{2}}{W(s)} \right\} (t) \cdot \mathcal{L}^{-1} \left\{ \frac{  u(s)(s+\Gamma_{N}/2)}{W(s)} \right\} (t) \right. \nonumber \\
&+&V^{2}_{12}(n_{2\uparrow}(0)+n_{2\downarrow}(0)-1)\mathcal{L}^{-1} \left\{ \frac{ s+\Gamma_{N}/2}{W(s)} \right\} (t) \cdot \mathcal{L}^{-1} \left\{ \frac{  u(s)}{W(s)} \right\} (t) \nonumber \\
&-&\left. \frac{\Gamma_{N}V^{2}_{12}}{2\pi}\int_{-\infty}^{\infty}d\varepsilon(1-2f_{N}(\varepsilon))\mathcal{L}^{-1} \left\{ \frac{ u(s)}{(s+i\varepsilon)W(s)}  \right\}(t) \cdot\mathcal{L}^{-1} \left\{ \frac{s+\Gamma_{N}/2}{(s-i\varepsilon)W(s)}  \right\}(t) \right],
\label{eq:B8}
\end{eqnarray}
\begin{eqnarray}
&&\langle \hat{c}_{2\downarrow}(t)\hat{c}_{2\uparrow}(t)\rangle = iV^{2}_{12}\frac{\Gamma_{S}}{2} \left[ (1-n_{1\uparrow}(0)-n_{2\uparrow}(0))\mathcal{L}^{-1} \left\{ \frac{ u(s)}{W(s)} \right\} (t) \cdot \mathcal{L}^{-1} \left\{ \frac{s+\Gamma_{N}/2}{W(s)} \right\} (t) \right. \nonumber \\
&+&\left[ 1-\sum_{\sigma}n_{2\sigma}(0) \right] \mathcal{L}^{-1} \left\{ \frac{ 1}{W(s)} \right\} (t) \cdot \ \mathcal{L}^{-1} \left\{ \frac{ u(s)V_{12}^{2}}{W(s)(s+\Gamma_{N}/2)}  -\frac{1}{s+\Gamma_{N}/2}\right\} (t) 
+\frac{\Gamma_{N}}{2\pi}\int_{-\infty}^{\infty}d\varepsilon(1-2f_{N}(\varepsilon))
\nonumber\\
&\cdot&\mathcal{L}^{-1} \left\{ \frac{V^{2}_{12} u(s)}{(s+i\varepsilon)(s+\Gamma_{N}/2)W(s)}- \frac{1}{(s+i\varepsilon)(s+\Gamma_{N}/2)}  \right\}(t) \cdot \left.  \mathcal{L}^{-1} \left\{ \frac{1}{(s-i\varepsilon)W(s)}  \right\}(t) \right] ,
\label{eq:B9}
\end{eqnarray}
\begin{eqnarray}
&&\langle \hat{c}_{1\downarrow}(t)\hat{c}_{2\uparrow}(t)\rangle = V_{12}\frac{\Gamma_{S}}{2}\left[ n_{1\uparrow}(0) \mathcal{L}^{-1} \left\{ \frac{(s+\Gamma_{N}/2)^{2}}{W(s)} \right\} (t) \cdot\mathcal{L}^{-1} \left\{ \frac{ u(s)}{W(s)} \right\} (t) 
+ n_{2\uparrow}(0) \mathcal{L}^{-1} \left\{ \frac{s+\Gamma_{N}/2}{W(s)} \right\} (t)
\right. \nonumber\\
&\cdot &  \mathcal{L}^{-1} \left\{ \frac{V^{2}_{12} u(s)}{(s+\Gamma_{N}/2)W(s)}- \frac{1}{(s+\Gamma_{N}/2)}  \right\} (t)
- (1-n_{1\downarrow}(0)) \mathcal{L}^{-1} \left\{ \frac{(s+\Gamma_{N}/2) u(s)}{W(s)} \right\} (t) \cdot \mathcal{L}^{-1} \left\{ \frac{s+\Gamma_{N}/2}{W(s)} \right\} (t)
 \nonumber\\
&-&  V^{2}_{12}(1-n_{2\downarrow}(0)) \mathcal{L}^{-1} \left\{ \frac{ u(s)}{W(s)} \right\} (t) \cdot \mathcal{L}^{-1} \left\{ \frac{1}{W(s)} \right\} (t) - \frac{\Gamma_{N}V^{2}_{12}}{2\pi}\int_{-\infty}^{\infty}d\varepsilon(1-f_{N}(\varepsilon)) \mathcal{L}^{-1} \left\{ \frac{ u(s)}{(s+i\varepsilon)W(s)}  \right\}(t)\nonumber\\
&\cdot & \mathcal{L}^{-1} \left\{ \frac{1}{(s-i\varepsilon)W(s)}  \right\}(t) 
+ \frac{\Gamma_{N}}{2\pi}\int_{-\infty}^{\infty}d\varepsilon f_{N}(\varepsilon) \mathcal{L}^{-1} \left\{ \frac{s+\Gamma_{N}}{(s-i\varepsilon)W(s)}  \right\}(t) \cdot \mathcal{L}^{-1} \left\{ \frac{V^{2}_{12} u(s)}{(s+i\varepsilon)(s+\Gamma_{N}/2)W(s)} \right.
\nonumber\\
& - & \left. \left.  \frac{1}{(s+i\varepsilon)(s+\Gamma_{N}/2)}  \right\}(t)\right],
\label{eq:B10}
\end{eqnarray}
\twocolumngrid
\noindent
with the auxiliary function $u(s)$ and $W(s)$ defined in Eqs.~(\ref{eq:9},\ref{eq:10}). 
For the DQD-S case, $\Gamma_{N}=0$ (Appendix \ref{sec.dynamics}), the nonvanishing values refer only to the imaginary (real) part of the on-dot $i=j$ (inter-dot $i \neq j$) pairings. The additional coupling of QD$_{2}$ to the normal lead allows the system to relax, evolving to its asymptotic (stationary) limit through a series of damped quantum oscillations. In consequence, the oscillating imaginary parts of $\langle c_{j\downarrow}(t)c_{j\uparrow}(t)\rangle$ and the real part of $\langle c_{1\downarrow}(t)c_{2\uparrow}(t)\rangle$ are now bounded between the curves $\pm \frac{\Gamma_{S}}{2}\exp(-\Gamma_{N}t/2)$ and $\pm \frac{\Gamma_{S}}{4}\exp(-\Gamma_{N}t/2)$. In contrast to the case of $\Gamma_{N}=0$, the real parts of both on-dot pairing functions are finite and they smoothly evolve from zero to their steady limit values.
Similar tendency can be observed for the imaginary part of the inter-dot pairing function, whose asymptotic value is rather residual. 

\begin{figure}
	\centerline{\includegraphics[width=0.9\linewidth]{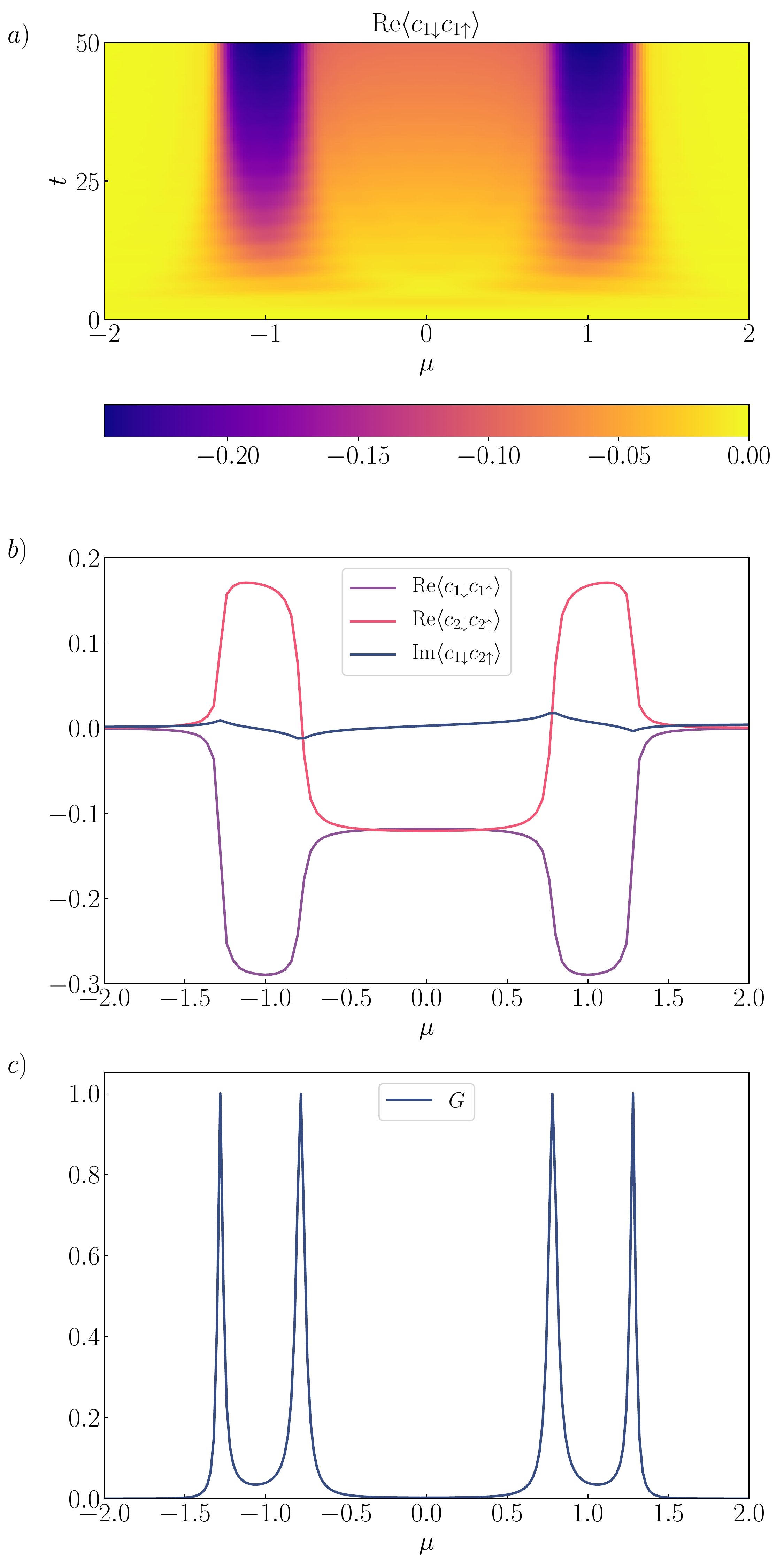}}
	\caption{(a) The real part of $\left< \hat{c}_{1\downarrow}(t) \hat{c}_{1\uparrow}(t)\right>$ as function of the bias voltage $\mu$ and time.
		(b) The asymptotic ($t\rightarrow\infty$) values  of Re$\left< \hat{c}_{1\downarrow}(t) \hat{c}_{1\uparrow}(t)\right>$,  Re$\left< \hat{c}_{2\downarrow}(t) \hat{c}_{2\uparrow}(t)\right>$ and  Im$\left< \hat{c}_{1\downarrow}(t) \hat{c}_{2\uparrow}(t)\right>$. (c) The stationary limit value of the differential conductance as a function of $\mu$. Results are obtained for $\Gamma_{N}=0.1$, $ \Gamma_{S}=1$, $V_{12}=1$, $\varepsilon_{j\sigma}=0$.}
	\label{fig9}
\end{figure}

Figure \ref{fig9} shows the real part of $\langle \hat{c}_{1\downarrow}(t)\hat{c}_{1\uparrow}(t)\rangle$ with respect to time and bias voltage (top panel), compared with the asymptotics of the real parts of $\lim_{t\rightarrow\infty}\langle \hat{c}_{1/2\downarrow}(t) \hat{c}_{1/2\uparrow}(t) \rangle$, the imaginary part of $\lim_{t\rightarrow\infty}\langle \hat{c}_{1\downarrow}(t) \hat{c}_{2\uparrow}(t) \rangle$ (middle panel) and the differential conductance (bottom panel). We can notice a coincidence between the positions of the excitation energies appearing in the differential conductance (bottom panel) with the characteristic features manifested in the pairing functions. Namely, the real parts of the on-dot pairing functions (which are strongly energy-dependent) have the inflexion points exactly at quasiparticle energies of the in-gap bound states (top and middle panels). On the other hand, the imaginary part of the inter-dot pairing function exhibits a jump of its derivative $\frac{\partial}{\partial \mu}\langle \hat{c}_{1\downarrow}(t)\hat{c}_{2\uparrow}(t)\rangle$ from the positive (negative) to negative (positive) values. It occurs exactly at bias voltages equal to the bound states energies.
Formally, these characteristic features originate from common poles of the diagonal and off-diagonal parts of the Green's function in the particle-hole (Nambu) representation.

\section{Coulomb repulsion effects}
\label{sec:Coulomb}

In realistic systems the repulsive on-dot interactions $U_{j}\hat{n}_{j\uparrow}\hat{n}_{j\downarrow}$
would compete with the proximity-induced electron pairing,
affecting the subgap bound states. Under stationary conditions this issue has been
investigated by various methods (see e.g.\ Ref.\ \cite{Rodero-11} for a survey).
In particular, the considerations of the DQD horizontally  embedded between
either normal and superconducting leads \cite{Tanaka.2010} or two superconductors \cite{Estrada_Saldana.2020}
have indeed shown a remarkable influence of the correlation effects.
To our knowledge, however, the transient dynamics of the correlated quantum dots
in these nanostructures has not been studied yet.
In this section we briefly address such problem.

The essential features due to the quench dynamics
of a correlated single quantum dot placed in the superconducting nanojunctions
has been previously explored in a perturbative framework \cite{LevyYeyati-2017}.
Perturbative approach, formulated in an appropriate way,
could qualitatively reproduce the results of such sophisticated methods
as NRG-type calculations \cite{Souto-2018}.
This fact encouraged us to perform the lowest-order perturbative analysis
for the same set of model parameters as used in Ref.\ \cite{Tanaka.2010},
focusing on the symmetric case, $\varepsilon_{j\sigma}=-U/2$ ($U\equiv U_{1}=U_{2}$),
where the Coulomb repulsion is most efficient.
For $U<2\Gamma_{S}$, $\Gamma_{N}=\Gamma_{S}$ and  $V_{12}/\Gamma_{N}=0.5-2$,
we have computed the linear conductance as a function of $V_{12}$,
qualitatively reproducing the NRG results \cite{Tanaka.2010}.
Obviously our mean field study (\ref{eq:HFB_1}) is reliable only in the weak interaction case,
$U<\Gamma_{\beta}$. In particular, for $U<\Gamma_{S}$,
the system is dominated by the Andreev scattering,
whereas for $U>2\Gamma_{S}$ the Kondo physics plays a major role \cite{Tanaka.2010,Estrada_Saldana.2020}.

To treat the correlations effects, we first make use of the 
Hartree-Fock-Bogoliubov (HFB) decoupling scheme
\begin{eqnarray}
\hat{n}_{j\uparrow}\hat{n}_{j\downarrow} &\simeq & \hat{n}_{j\uparrow}\; n_{j\downarrow}(t)
+ \hat{n}_{j\downarrow} \; n_{j\uparrow}(t)
\label{eq:HFB_1}\\
&+&\hat{c}_{j\uparrow}^{\dagger}\hat{c}_{j\downarrow}^{\dagger}\langle\hat{c}_{j\downarrow}\hat{c}_{j\uparrow}\rangle
+\hat{c}_{j\downarrow}\hat{c}_{j\uparrow}\langle\hat{c}_{j\uparrow}^{\dagger}\hat{c}_{j\downarrow}^{\dagger}\rangle ,
\nonumber
\end{eqnarray}
which yields the renormalized energy levels  $\tilde{\varepsilon}_{j\sigma}(t)=\varepsilon_{j\sigma}+U_{j}n_{j\bar{\sigma}}(t)$,
where $\bar{\sigma}$ stands for the opposite spin to $\sigma$,
and important corrections to the time-dependent pairing potentials $\Delta_{1}(t)=\Gamma_{S}/2-U_{1}\langle c_{1\downarrow}(t)c_{1\uparrow}(t)\rangle$ and $\Delta_{2}(t)=-U_{2}\langle c_{2\downarrow}(t)c_{2\uparrow}(t)\rangle$.
Combining the interactions with the superconducting proximity effect can be effectively described by the following Hamiltonian
\begin{eqnarray}
\hat{H}^{HFB} \approx  \sum_{j,\sigma} \tilde{\varepsilon}_{j\sigma}(t) \hat{c}_{j\sigma}^{\dagger}\hat{c}_{j\sigma}
\!-\!\sum_{j} \left( \Delta_{j}(t) \hat{c}_{j\uparrow}^{\dagger} \hat{c}_{j\downarrow}^{\dagger} + \mbox{\rm h.c.} \right)
\label{effective_model} \\
+  \sum_{\sigma} \left[ \left( V_{12}\hat{c}_{1\sigma}^{\dagger} + \sum_{\textbf{k}}V_{N\textbf{k}}
\hat{c}_{N\textbf{k}\sigma}^{\dagger} \right) \hat{c}_{2\sigma} + \mbox{\rm h.c.} \right] + \hat{H}_{N}  ,
\nonumber
\end{eqnarray}
where the time-dependent energy levels $\tilde{\varepsilon}_{j\sigma}(t)$ and on-dot pairings $\Delta_{j}(t)$ must be determined numerically.
We have self-consistently computed the time-dependent $n_{j \sigma}(t)$,  $\langle \hat{c}_{j\downarrow}(t)\hat{c}_{j\uparrow}(t) \rangle$,  the current $j_{N\sigma}(t)$, and its differential conductance $G_{\sigma}(\mu,t)$, using the procedure outlined by us in Ref.~\cite{Taranko-2018} (see Appendix B). For this purpose we have solved the differential equations of motion for $n_{j\sigma}(t)$, $\langle \hat{c}_{j\downarrow}(t)\hat{c}_{j\uparrow}(t)\rangle$ and $\langle \hat{c}_{1\sigma}^{\dagger}(t)\hat{c}_{2\sigma '}(t)\rangle$
at intermediate steps computing also the correlation functions $\langle \hat{c}_{1\sigma}^{\dagger}(t)\hat{c}_{N\textbf{k}\sigma}(0)\rangle$,  $\langle \hat{c}_{1\sigma}(t)\hat{c}_{N\textbf{k}\bar{\sigma}}(0)\rangle$,  $\langle \hat{c}_{2\sigma}^{\dagger}(t)\hat{c}_{N\textbf{k}\sigma}(0)\rangle$ and  $\langle \hat{c}_{2\sigma}(t)\hat{c}_{N\textbf{k}\bar{\sigma}}(0)\rangle$. We have calculated these quantities iteratively within the Runge-Kutta algorithm, starting from their initial  ($t=0$) values.

\begin{figure*}[ht]
\centerline{\includegraphics[width=1\linewidth]{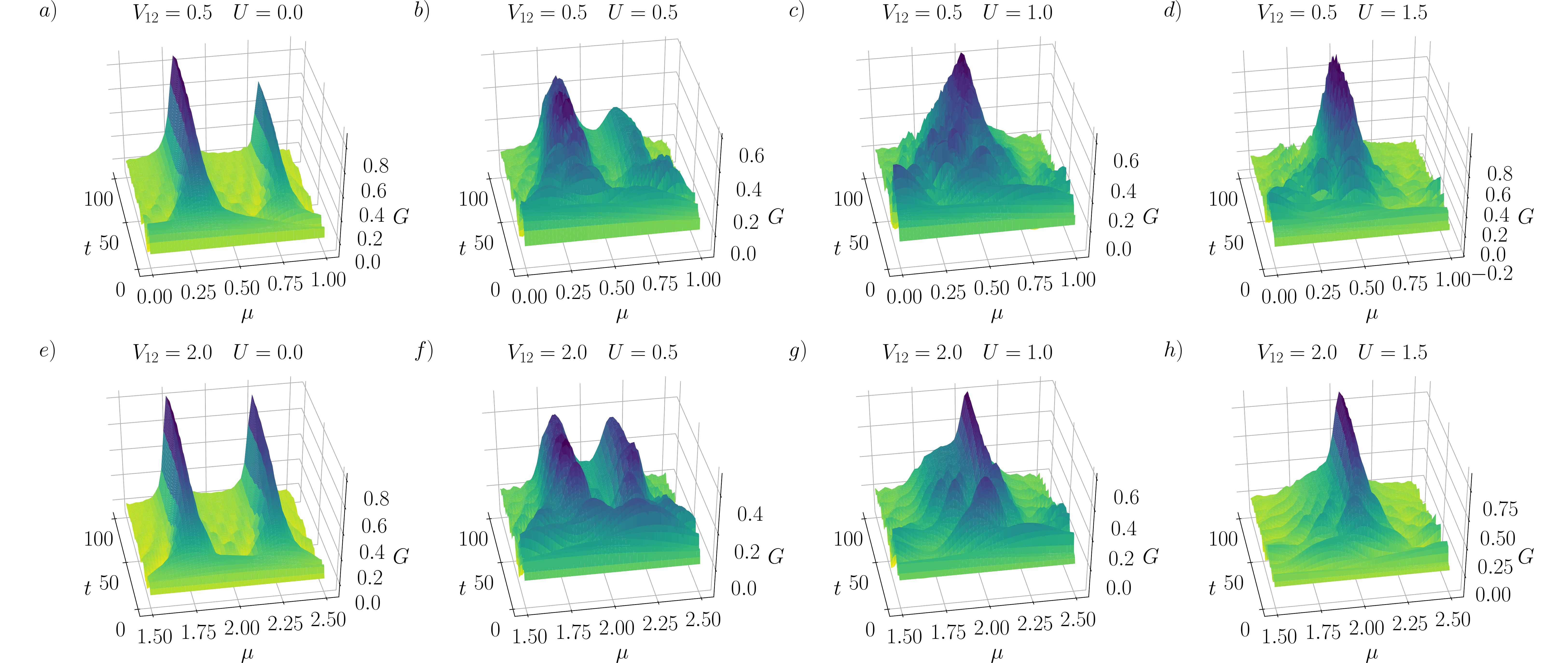}}
\caption{The time-dependent differential conductance $G(\mu,t)$ [in units of $2e^{2}/h$] as a function the bias voltage $\mu$ obtained in the weak $V_{12}=0.5$ (upper panels) and strong interdot coupling limit $V_{12}=2$ (bottom panels) for several values of the Coulomb potential (as indicated), assuming  $\varepsilon_{j\sigma}=-U/2$, $\Gamma_{N}=0.1$ and $\Gamma_{S}\equiv 1$.}
\label{fig10_MF}
\end{figure*}

Figure \ref{fig10_MF} displays the typical evolution of the differential conductance
obtained for several values of the Coulomb potential, assuming small, $V_{12}=0.5$, and large, $V_{12}=2$, interdot couplings.
As the Andreev conductance is symmetric with respect to the bias voltage, $G(\mu,t)=G(-\mu,t)$,
we show its variation only for the positive bias $\mu$ where all dynamical features can be well recognized.
Upon increasing the Coulomb repulsion $U$ ($U_1=U_2\equiv U$),
the two-peak structure (characteristic for the noninteracting system)
undergoes a gradual reconstruction into a single broad peak.
This tendency indicates that the Coulomb repulsion suppresses the effects caused
by both the interdot hybridization and the superconducting proximity effect.
The time needed for the development of such final structure
(observable in the differential conductance with respect to the bias voltage $\mu$)
turns out to be $\sim 100 \hbar/\Gamma_{S}$.
For the experimentally realistic coupling $\Gamma_{S}\sim 200$ $\mu$eV,
this characteristic time-scale would be $0.3-0.4$ $\mu$sec.

\begin{figure*}[ht]
	\includegraphics[width=1\linewidth]{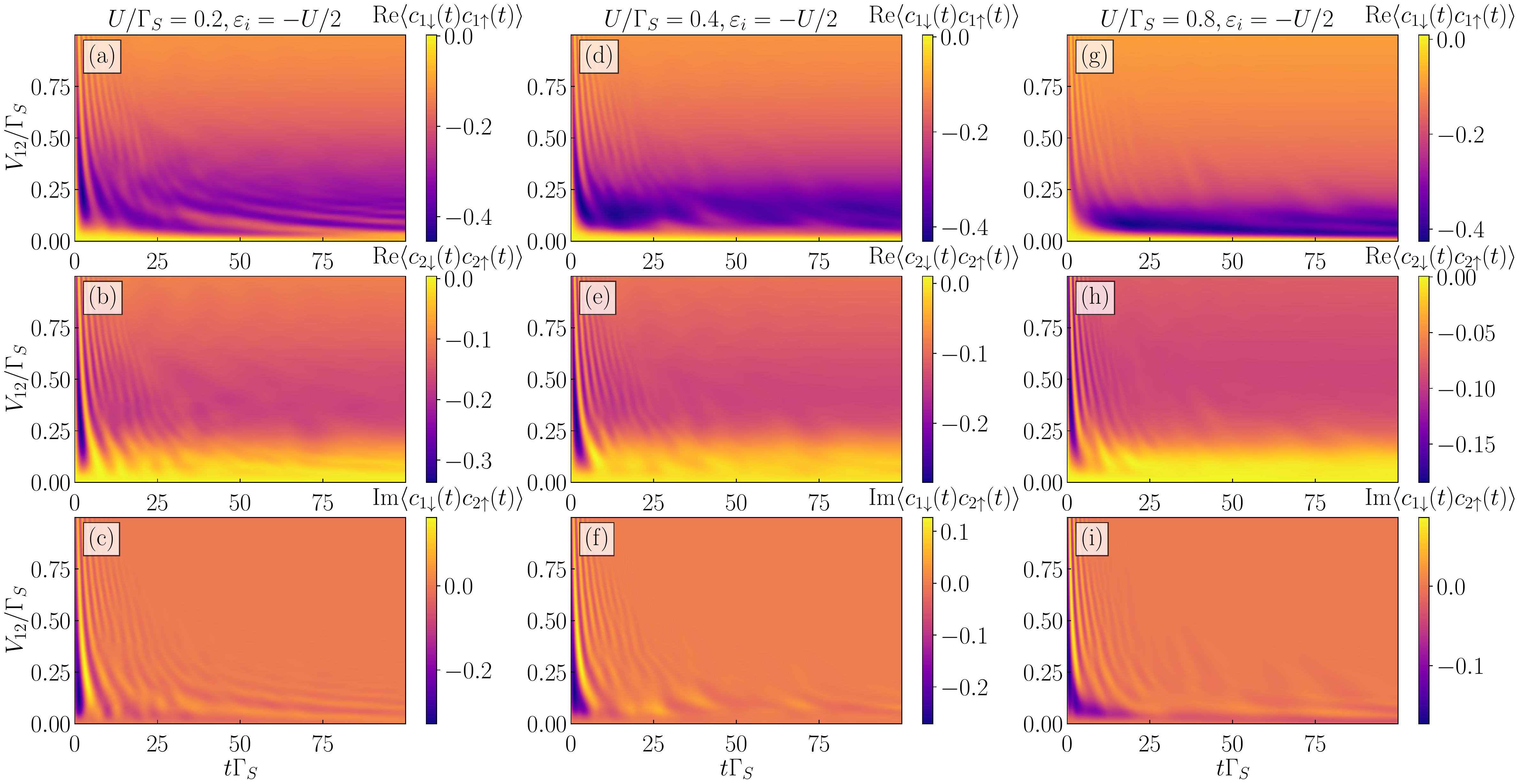}
	\caption{The real part of the on-dot pairings $\left< \hat{c}_{j\downarrow}(t) \hat{c}_{j\uparrow}(t)\right>$ (top and middle panels)
		and imaginary part of the inter-dot pairing $\left< \hat{c}_{1\downarrow}(t) \hat{c}_{2\uparrow}(t)\right>$ (bottom panel) with respect to time (horizontal axis) and the coupling $V_{12}$ (vertical axis). Results are obtained by tNRG calculations
		for several values of the Coulomb potential (as indicated), assuming the half-filled quantum dots $\varepsilon_{j}=-U/2$ and $\Gamma_{N}/\Gamma_{S}=0.1$.}
	\label{fig11_tNRG}
\end{figure*}

For more credible determination of the correlation effects
beyond the perturbative framework, we have additionally
used the time-dependent numerical renormalization group technique \cite{Wilson1975,Anders2005,Anders2006,Bulla2008,Costi2014,Costi2014generalization,NRG_code}.
This approach allows for treating correlations in very accurate manner,
however, it is restricted to unbiased junctions.
The tNRG employs the Wilson's numerical renormalization group (NRG)
method to solve the initial ($\hat{H}_0$) and final ($\hat{H}$) Hamiltonians
essential to evaluate the quench dynamics according to the general form of time-dependent Hamiltonian
\begin{equation}\label{Eq:Hamiltonian_TD}
  \hat{H}(t) = \theta(-t)\hat{H_0} + \theta(t)\hat{H}.
\end{equation}
The diagonalization of both Hamiltonians is performed
in $N$ iterations with $N_K$ energetically lowest-lying eigenstates retained at each iteration.
These kept eigenstates, tagged with superscript $K$,
are used in consecutive iterations to build new product states
corresponding to the addition of another site of the Wilson chain.
The remaining states are referred to as discarded,
as well as all states from the last iteration of the procedure, and are tagged with superscript $D$.
All discarded states of the corresponding Hamiltonians are used to
span the full many-body initial and final eigenbases \cite{Anders2005}
\begin{equation} \label{Eq:completeness}
\sum_{nse}|nse\rangle^{\!D}_{0} \,{}^{D}_{\,0}\!\langle nse| \!=\! \mathbbm{\hat{1}} \;\;\;\,
 {\rm and}
 \,\;\;\; \sum_{nse}|nse\rangle^{\!D} \,{}^D \!\langle nse| \!=\! \mathbbm{\hat{1}}.
\end{equation}
Here, the index $s$ refers to the eigenstates obtained at $n$-th iteration
 and the index $e$ expresses the environmental part of the Wilson chain.
Due to the energy-scale separation, these eigenstates are good approximations
of the eigenstates of the full NRG Hamiltonians.

We have computed the dynamical quantities of the unbiased N-DQD-S heterostructure,
determining the expectation values of the observables in frequency domain
$O(\omega)\equiv\langle \mathcal{\hat{O}}(\omega) \rangle$.
The formula for $O(\omega)$ in terms of the designated eigenstates can be written as
\begin{eqnarray}\label{Eq:Ow}
    O(\omega) &=&\!\!
       \sum_{n}^{ XX'\neq KK}\sum_{n'} \sum_{ss'e}  {}^{X}\! \langle nse|w_{n'} \hat{\rho}_{0n'}| ns'e\rangle^{\! X'} \nonumber\\
   &\times& {}^{ X'}\! \langle ns'e|\mathcal{\hat{O}}|nse\rangle^{\! X} \; \delta(\omega + E_{ns}^{X} - E_{ns'}^{X'}).
\end{eqnarray}
Here, $\hat{\rho}_{0n'}$ denotes the contribution to the initial density matrix
from the $n'$-th iteration and $w_{n'}$ is the corresponding weight
after tracing out the environmental states,
while the initial full density matrix $\hat{\rho}_0$ built from $\hat{H}_0$ at thermal equilibrium reads \cite{Andreas_broadening2007}
\begin{equation}
\hat{\rho}_0=\sum_{nse}\frac{e^{-\beta E_{0ns}^D}}{Z} |nse\rangle^{\!D}_{0} \,{}^{D}_{\,0}\!\langle nse|,
\end{equation}
where $\beta \equiv (k_{B}T)^{-1}$ is the inverse temperature and
$Z\equiv\sum_{nse} e^{-\beta E_{0ns}^D}$ is the partition function.

In the following steps, the obtained collection of Dirac delta peaks with corresponding weights
is weakly smoothed with a log-Gaussian function and broadening parameter $b \leq 0.1$.
Finally, a Fourier-transformation back into the time domain is applied \cite{Andreas2012}
\begin{equation}
    O(t)=\int_{-\infty }^{\infty}O(\omega)e^{-i \omega t} d \omega.
\end{equation}
In performed tNRG calculations we have assumed the discretization parameter $\Lambda = 2$,
the length of the Wilson chain to consist of $N=100$ sites and we have kept $N_K=2000$ eigenstates at each iteration.
More detailed description of the tNRG implementation and technicalities has been discussed in Ref.~\cite{WrzesniewskiWeymann-2019}.

Figure \ref{fig11_tNRG} presents the real/imaginary parts of the electron pairing induced on individual quantum dots
(the upper and middle panels) and between them (the bottom panels) for three different values of Coulomb potential: $U/\Gamma_{S}=0.2$,
$0.4$, and $0.8$. Here, we annotate that for tNRG results we have evaluated the quench
exclusively in the coupling to the superconducting lead $\Gamma_S$.
Other couplings are assumed to be time-independent and have values as specified in Fig. \ref{fig11_tNRG}.
This modification of the quench protocol allowed us to remove weak
and non-relevant dynamics associated with switching on other couplings,
while the role of the superconducting correlations is now more evident.
We checked numerically that in both scenarios the results and conclusions are in agreement.

We clearly notice that repulsive interactions suppress the pairings of all these channels. Comparison
of these quantities against time at some fixed inter-dot coupling (for instance $V_{12}=0.5$) indicates
that the quantum oscillations become faster upon increasing the Coulomb potential $U$. This speed-up of quantum oscillations
stems from renormalization of the in-gap states energies observable also in the mean-field calculations (Fig.~\ref{fig10_MF}). Additionally, we notice that the region of transient effects gradually shrinks with
increasing the Coulomb potential. The latter effect can be indirectly assigned to suppression of the superconducting
proximity effect (recall that the quantum oscillations are here driven by Rabi-type transitions between pairs of
in-gap bound states).

The interdot coupling $V_{12}$ plays an important role in the distribution
of the on-dot pairing potential between the coupled quantum dots.
For relatively weak values, $V_{12}/\Gamma_S<0.25$,
the strong on-dot pairing potential is present on the quantum dot directly coupled to the superconductor,
while the second dot is almost unaffected by the proximity effect.
However, as the interdot coupling is amplified, it mediates the
superconducting correlations onto the second dot.
For values $V_{12}/\Gamma_S>0.5$, the on-dot pairing potential is more
evenly distributed between both dots.
This observation reveals crucial role of interdot coupling in transferring the superconducting correlations.
Another very important feature can be seen for the weak interdot coupling.
For all pairing channels we clearly notice blockade of the superconducting proximity effect,
strictly due to the initial single occupancy of both quantum dots.
This brings us to the important conclusion that dynamical signatures of the triplet/Andreev
blockade should be well observable in the correlated N-DQD-S nanostructures,
whenever the coupling between the quantum dots is weak.

Summarizing this section, we emphasize that a competition of the repulsive on-dot interactions with the superconducting
proximity effect is evident, both in the stationary and dynamical properties. The magnitude of electron pairing induced
on each quantum dot and between them is considerably suppressed by the interactions. Furthermore, the quantum oscillations
become faster and transient phenomena survive over some narrower time-region upon increasing the Coulomb potential. 

\section{Summary and outlook}
\label{sec:summary}

We have investigated the dynamical effects observable in the double quantum dot (DQD) abruptly embedded between the superconducting and metallic leads. Transient phenomena of the uncorrelated setup have been explored by solving the coupled equations of motion, treating the initial constraints within the Laplace transform approach. Focusing on the subgap regime, we have derived analytical expressions for the charge occupancy of both quantum dots, the induced on-dot and inter-dot electron pairings, and the  currents flowing between neighboring constituents of N-DQD-S heterostructure.
The time-dependent quantities (except the differential conductance) have been represented by contributions, dependent on the initial DQD fillings and on their couplings to the external leads. These expressions guided us to identify the characteristic time-scales of transient phenomena, manifested by:
(i) the Rabi-type quantum oscillations due to transitions between the pairs of in-gap bound states and
(ii) the relaxation processes involving a continuous spectrum of the metallic lead.

To single out the quantum oscillations, we have analyzed them for DQD coupled only to the superconducting reservoir (Appendix \ref{sec.dynamics}). Under such circumstances all physical quantities would be periodic in time, unless the higher energy electronic states from outside the pairing gap were taken into consideration \cite{LevyYeyati-2017}. Our analytical expressions  (\ref{eq:xx43}-\ref{eq:xx45}) indicate  that the on-dot pairing functions are purely imaginary whereas the inter-dot pairing function is purely real. We have investigated the components of quantum oscillations in the strong $V_{12} >\Gamma_{S}$ and weak $V_{12} < \Gamma_{S}$ interdot couplings, respectively. Furthermore, we have also inspected under what circumstances the superconducting proximity effect is going to be blocked, preventing the Cooper pairs from leaking onto the quantum dots. 
 We have found that for the initial triplet configuration of the DQD-S system, the charge flow $j_{S\sigma}(t)$ between the superconducting lead and neighboring quantum dot is completely forbidden. 

In the N-DQD-S junctions similar blockade is still present, although in less severe version because electrons can flow back and forth to/from the normal lead. Under the stationary conditions such {\it triplet-blockade} has been reported experimentally in the Josephson (S-DQD-S) junction \cite{Paaske-2020} and its analogue, so called {\it Andreev-blockade}, has been recently
evidenced for N-DQD-S heterostructure \cite{Frolov-2021}. Suppression of the superconducting proximity effect occurs also in presence of the correlations, especially in the weak interdot coupling regime. Additionally, we have shown that the time-resolved Andreev conductance  can probe a buildup of the in-gap bound states and indirectly detect the dynamical superconducting proximity effect.

In future it would be worthwhile to study transient phenomena of the interacting quantum dots, focusing on the parity crossings and realization of the subgap Kondo effect. We hope that our analytical results obtained for the noninteracting system could serve as a useful benchmark for such project. Another challenging issue can be related to the Majorana-type versions of the in-gap bound states \cite{Prada-2020} with appealing perspectives to use them in semiconductor-based superconducting qubits and quantum computing \cite{Aguado-2020}.

\begin{acknowledgments}
This work was supported by the National Science Centre (NCN, Poland) under
the grants UMO-2017/27/B/ST3/01911 (RT, BB), UMO-2018/29/N/ST3/01038 (KW), 
and UMO-2018/29/B/ST3/00937 (IW, TD).
\end{acknowledgments}

\appendix
\section{Laplace transforms}
\label{app:A}


We derive here the Laplace transforms for $\hat{c}_{j\sigma}(s)$ and $\hat{c}_{S {\textbf{q}}/N {\textbf{k}}\sigma}(s)$ required for
the determination of the time-dependent physical quantities discussed in this paper.
Upon transforming the Heisenberg equations we obtain
\onecolumngrid
\begin{eqnarray}
(s+i\varepsilon_{1\uparrow})\hat{c}_{1\uparrow}(s)&=&-i\sum _{ {\textbf{q}}}V _{S {\textbf{q}}}\hat{c}_{S {\textbf{q}}\uparrow}(s)-iV_{12}\hat{c}_{2\uparrow}(s)+\hat{c}_{1\uparrow}(0),
\label{A1}
\\
(s-i\varepsilon_{1\downarrow})\hat{c}_{1\downarrow}^{\dagger}(s)&=&i\sum _{ {\textbf{q}}}V _{S {\textbf{q}}}\hat{c}_{S {\textbf{q}}\downarrow}^{\dagger}(s)+iV_{12}\hat{c}_{2\downarrow}^{\dagger}(s)+\hat{c}_{1\downarrow}^{\dagger}(0),
\label{A2}
\\
(s+i\varepsilon _{S {\textbf{q}}})\hat{c}_{S {\textbf{q}}\uparrow}(s)&=&-iV _{S {\textbf{q}}}\hat{c}_{1\uparrow}(s)-i\Delta \hat{c}_{S- {\textbf{q}}\downarrow}^{\dagger}(s)+\hat{c}_{S {\textbf{q}}\uparrow}(0),
\label{A3}
\\
(s-i\varepsilon _{S {\textbf{q}}})\hat{c}_{S- {\textbf{q}}\downarrow}^{\dagger}(s)&=&iV _{S {\textbf{q}}}\hat{c}_{1\downarrow}^{\dagger}(s)-i\Delta \hat{c}_{S {\textbf{q}}\uparrow}(s)+c_{S- {\textbf{q}}\downarrow}^{\dagger}(0),
\label{A4}
\end{eqnarray}
and
\begin{eqnarray}
(s+i\varepsilon_{2\uparrow})\hat{c}_{2\uparrow}(s)&=&-i\sum _{ {\textbf{k}}}V _{N {\textbf{k}}}\hat{c}_{N {\textbf{k}}\uparrow}(s)-iV_{12}\hat{c}_{1\uparrow}(s)+\hat{c}_{2\uparrow}(0),
\label{A5}
\\
(s-i\varepsilon_{2\downarrow})\hat{c}_{2\downarrow}^{\dagger}(s)&=&i\sum _{ {\textbf{k}}}V _{N {\textbf{k}}}\hat{c}_{N {\textbf{k}}\downarrow}^{\dagger}(s)+iV_{12}\hat{c}_{1\downarrow}^{\dagger}(s)+\hat{c}_{2\downarrow}^{\dagger}(0),
\label{A6}
\\
(s+i\varepsilon _{N {\textbf{k}}})\hat{c}_{N {\textbf{k}}\uparrow}(s)&=&-iV _{N {\textbf{k}}}\hat{c}_{2\uparrow}(s)+\hat{c}_{N {\textbf{k}}\uparrow}(0),
\label{A7}
\\
(s-i\varepsilon _{N {\textbf{k}}})\hat{c}_{N {\textbf{k}}\downarrow}^{\dagger}(s)&=&iV _{N {\textbf{k}}}\hat{c}_{2\downarrow}^{\dagger}(s)+\hat{c}_{N {\textbf{k}}\downarrow}^{\dagger}(0) .
\label{A8}
\end{eqnarray}
Eqs.\ (\ref{A1}-\ref{A4}) are coupled to (\ref{A5}-\ref{A8}) through the inter-dot coupling $V_{12}$.
After some lengthy but straightforward algebra, we can simplify them to the following compact form
\begin{eqnarray}
\left( s+i\varepsilon_{S {\textbf{q}}\uparrow}+\sum _{ {\textbf{q}}}\frac{V _{S {\textbf{q}}}^{2}(s-i\varepsilon _{S {\textbf{q}}})}{s^{2}+\varepsilon _{S {\textbf{q}}}^{2}+\Delta^{2}}\right) \hat{c}_{1\uparrow}(s) &=&
-\sum _{ {\textbf{q}}}\frac{V _{S {\textbf{q}}}^{2}\Delta}{s^{2}+\varepsilon _{S {\textbf{q}}}^{2}+\Delta^{2}}\hat{c}_{1\downarrow}^{\dagger}(s)-iV_{12}\hat{c}_{2\uparrow}(s)+\hat{a}_{1},
\label{A12}\\
\left( s-i\varepsilon_{1\downarrow}+\sum _{ {\textbf{q}}}\frac{V _{S {\textbf{q}}}^{2}(s+i\varepsilon _{S {\textbf{q}}})}{s^{2}+\varepsilon _{S {\textbf{q}}}^{2}+\Delta^{2}}\right) \hat{c}_{1\downarrow}^{\dagger}(s) &=&
-i\sum _{ {\textbf{q}}}\frac{V _{S {\textbf{q}}}^{2}\Delta}{s^{2}+\varepsilon _{S {\textbf{q}}}^{2}+\Delta^{2}}\hat{c}_{1\uparrow}(s)+iV_{12}\hat{c}_{2\downarrow}^{\dagger}(0)+\hat{a}_{2},
\label{A14}
\\
\left( s+i\varepsilon_{2\uparrow}+\sum _{ {\textbf{k}}}\frac{V _{N {\textbf{k}}}^{2}}{s+i\varepsilon _{N {\textbf{k}}}} \right) \hat{c}_{2\uparrow}(s)&=&-iV_{12}\hat{c}_{1\uparrow}(s)+\hat{a}_{3},
\label{A15}
\\
\left( s-i\varepsilon_{2\uparrow}+\sum _{ {\textbf{k}}}\frac{V _{N {\textbf{k}}}^{2}}{s-i\varepsilon _{N {\textbf{k}}}}\right) \hat{c}_{2\downarrow}^{\dagger}(s)&=&iV_{12}\hat{c}_{1\downarrow}(s)+\hat{a}_{4},
\label{A16}
\end{eqnarray}
where the last components are defined as
\begin{eqnarray}
\hat{a}_{1}&=&\hat{c}_{1\uparrow}(0)-\sum _{ {\textbf{q}}}\frac{V _{S {\textbf{q}}}\Delta}{s^{2}+\varepsilon _{S {\textbf{q}}}^{2}+\Delta^{2}}\hat{c}_{S- {\textbf{q}}\downarrow}^{\dagger}(0) - i\sum _{ {\textbf{q}}}\frac{V _{S {\textbf{q}}}(s-i\varepsilon _{S {\textbf{q}}})}{s^{2}+\varepsilon _{S {\textbf{q}}}^{2}+\Delta^{2}}\hat{c}_{S {\textbf{q}}\uparrow}(0),
\label{A13} \\
\hat{a}_{2}&=&\hat{c}_{1\downarrow}^{\dagger}(0)+\sum _{ {\textbf{q}}}\frac{V _{S {\textbf{q}}}\Delta}{s^{2}+\varepsilon _{S {\textbf{q}}}^{2}+\Delta^{2}}\hat{c}_{S- {\textbf{q}}\uparrow}(0) + i\sum _{ {\textbf{q}}}\frac{V _{S {\textbf{q}}}(s+i\varepsilon _{S {\textbf{q}}})}{s^{2}+\varepsilon _{S {\textbf{q}}}^{2}+\Delta^{2}}\hat{c}_{S {\textbf{q}}\downarrow}^{\dagger}(0),
\label{A17}
\\
\hat{a}_{3}&=&-i\sum _{ {\textbf{k}}}\frac{V _{N {\textbf{k}}}}{s+i\varepsilon _{N {\textbf{k}}}}\hat{c}_{N {\textbf{k}}\uparrow}(0)+\hat{c}_{2\uparrow}(0),
\label{A18}
\\
\hat{a}_{4}&=&i\sum _{ {\textbf{k}}}\frac{V _{N {\textbf{k}}}}{s-i\varepsilon _{N {\textbf{k}}}}\hat{c}_{N {\textbf{k}}\downarrow}^{\dagger}(0)+\hat{c}_{2\downarrow}^{\dagger}(0).
\label{A19}
\end{eqnarray}
In the wide bandwidth limit
we can perform summations over momenta ${\textbf{k}}$ and $ {\textbf{q}}$ of the itinerant electrons. In this way we obtain the set of coupled equations (\ref{eq:3}-\ref{eq:6}) presented in the main part of this manuscript.

\section{Interdot charge flow and supercurrent}
\label{app:currents}

Here we provide detailed expressions for the interdot current $j_{12\sigma}(t)$ and the current $j_{S\sigma}(t)$
between QD$_{1}$ and superconductor. The charge flow $j_{12\sigma}(t)$ between the quantum dots can be calculated
from
\begin{equation}
\begin{aligned}
j_{12\sigma}(t)=-2V_{12}\textrm{Im}\langle \hat{c}^{\dagger}_{1\sigma}(t)\hat{c}_{2\sigma}(t)\rangle.
\end{aligned}
\label{eq:B3}
\end{equation}
Using the inverse Laplace transforms of $c_{j\sigma}(s)$ operators [Eqs.~(\ref{eq:7},\ref{eq:8})] we obtain
\begin{equation}
\begin{aligned}
j_{12\uparrow/\downarrow}(t)&=-2V^{2}_{12}\textrm{Re} \left[ n_{1\uparrow/\downarrow}(0)\mathcal{L}^{-1} \left\{ \frac{u(s)(s+\Gamma_{N}/2)}{W(s)}  \right\} (t)\cdot\mathcal{L}^{-1} \left\{ \frac{ u(s)}{W(s)}  \right\} (t)-n_{2\uparrow/\downarrow}(0)\mathcal{L}^{-1} \left\{ \frac{ u(s)}{W(s)}  \right\} (t) \right. \\
&\times  \mathcal{L}^{-1} \left\{ \frac{1}{s+\Gamma_{N}/2}(1-\frac{ u(s)V_{12}^{2}}{W(s)} ) \right\} (t)
+(1-n_{1\downarrow/\uparrow}(0))\Gamma_{S}^{2}/4  \mathcal{L}^{-1} \left\{ \frac{ (s+\Gamma_{N}/2)^{2}}{W(s)} \right\} (t) \cdot \mathcal{L}^{-1} \left\{ \frac{ s+\Gamma_{N}/2}{W(s)} \right\} (t) \\
&+V_{12}^{2}\Gamma_{S}^{2}/4(1-n_{2\downarrow/\uparrow}(0))  \mathcal{L}^{-1} \left\{ \frac{s+\Gamma_{N}/2}{W(s)} \right\} (t)\cdot \mathcal{L}^{-1} \left\{ \frac{1}{W(s)} \right\} (t) \\
&+\Gamma_{N}\Gamma_{S}^{2}V^{2}_{12}/8\pi\int_{-\infty}^{\infty}d\varepsilon(1-f_{N}(\varepsilon))\mathcal{L}^{-1} \left\{ \frac{s+\Gamma_{N}/2}{(s+i\varepsilon)W(s)}  \right\}(t) \cdot\mathcal{L}^{-1} \left\{ \frac{1}{(s-i\varepsilon)W(s)}  \right\}(t)\\
&\left. +\frac{\Gamma_{N}}{\pi}\int_{-\infty}^{\infty}d\varepsilon f_{N}(\varepsilon)\mathcal{L}^{-1} \left\{ \frac{ u(s)}{W(s)(s-i\varepsilon)}\right\}(t)\cdot  \mathcal{L}^{-1} \left\{ (\frac{V_{12}^{2} u(s)}{W(s)}-1)(\frac{1}{(s+i\varepsilon)(s+\Gamma_{N}/2)})  \right\}(t) \right].
\end{aligned}
\label{eq:B4}
\end{equation}
In a similar way, the current flowing from the superconducting lead to the first quantum dot is given by
\begin{equation}
\begin{aligned}
j_{S\sigma}(t)=2\textrm{Im}\left[ \sum _{ {\textbf{q}}}V _{S {\textbf{q}}} \langle \hat{c}^{\dagger}_{1\sigma}(t)\hat{c} _{S {\textbf{q}}\sigma}(t)\rangle \right],
\end{aligned}
\label{eq:B5}
\end{equation}
where $\hat{c}^{\dagger}_{1\sigma}(t)$ should be taken from the inverse Laplace transform of the Hermitian conjugation of (\ref{eq:7}). The inverse Laplace transform of $c _{S {\textbf{q}}\sigma}(s)$, calculated from Eqs.~(\ref{A1}-\ref{A8}), takes the following form
\begin{equation}
\begin{aligned}
\hat{c}_{S {\textbf{q}}\uparrow}(s)= \frac{1}{s^{2}+\varepsilon^{2} _{S {\textbf{q}}}+\Delta^{2}}\left[-i V _{S {\textbf{q}}}(s-i\varepsilon _{S {\textbf{q}}})\hat{c}_{1\uparrow}+V _{S {\textbf{q}}}\Delta \hat{c}^{\dagger}_{1\downarrow}(s) -i\Delta \hat{c}^{\dagger}_{S- {\textbf{q}}\downarrow}(0)+(s-i\varepsilon _{S {\textbf{q}}})\hat{c}_{S {\textbf{q}}\uparrow}(0) \right].
\end{aligned}
\label{eq:B6}
\end{equation}
In the limit $\Delta\rightarrow\infty$ we obtain
\begin{equation}
\begin{aligned}
j_{S\sigma}(t) &= \frac{\Gamma^{2}_{S}}{2} \textrm{Re} \left[ (1-n_{1\sigma}(0)-n_{1-\sigma}(0)) \mathcal{L}^{-1} \left\{ \frac{ (s+\Gamma_{N}/2)^{2}}{W(s)} \right\} (t) \cdot  \mathcal{L}^{-1} \left\{ \frac{ (s+\Gamma_{N}/2) u(s)}{W(s)} \right\} (t) \right. \\
&+ V^{2}_{12}(1-n_{2\sigma}(0)-n_{2-\sigma}(0))\mathcal{L}^{-1} \left\{ \frac{s+\Gamma_{N}/2}{W(s)} \right\} (t) \cdot  \mathcal{L}^{-1} \left\{ \frac{ u(s)}{W(s)} \right\} (t)\\
&\left. +\frac{\Gamma_{N}V^{2}_{12}}{2\pi}\int_{-\infty}^{\infty}d\varepsilon(1-2f_{N}(\varepsilon))\mathcal{L}^{-1} \left\{ \frac{s+\Gamma_{N}/2}{(s+i\varepsilon)W(s)}  \right\}(t) \cdot\mathcal{L}^{-1} \left\{ \frac{ u(s)}{(s-i\varepsilon)W(s)}  \right\}(t)  \right] .
\end{aligned}
\label{eq:B7}
\end{equation}
\twocolumngrid
\noindent
Since the supercurrent $j_{S\sigma}(t)$ originates from tunneling of electron pairs, therefore $j_{S\uparrow}(t)=j_{S\downarrow}(t)$.

\section{DQD coupled to superconductor}
\label{sec.dynamics}

Let us consider the case of $\Gamma_{N}=0$.
Under such circumstances, one can derive the analytical expressions for observables, which well illustrate
the dynamics induced by an abrupt coupling to the superconducting lead.

For $\Gamma_{N}=0$, Eqs.\ (\ref{eq:9},\ref{eq:10}) simplify to
\begin{eqnarray}
u(s)&=&s^{2}+V_{12}^{2},\\
W(s)&=&(s^{2}+V_{12}^{2})^{2}+\Gamma_{S}^{2}s^{2}/4  .
\label{eq:16}
\end{eqnarray}
The complex roots of (\ref{eq:8}) are given by $s_{1,2}=\pm i\bar{s}_{1}$ and $s_{3,4}=\pm i\bar{s}_{3}$,
where
\begin{equation}
\bar{s}_{1,3}=\frac{1}{2}\left(\sqrt{4V^{2}_{12}+\Gamma^{2}_{S}/4}\mp\frac{\Gamma_{S}}{2} \right) ,
\label{eq:17}
\end{equation}
thus the inverse Laplace transforms $\mathcal{L}^{-1}\{\hat{c}_{j\sigma}^{(\dagger)}(s)\}(t)$  can be obtained explicitly. In what follows,
we analyze the expectation values of various quantities, showing that they periodically oscillate in time with the characteristic frequencies.
In the absence of metallic lead ($\Gamma_{N}=0$),
the last two terms of Eqs.\ (\ref{eq:B1},\ref{eq:B2}) vanish, therefore $n_{j\sigma}(t)$ simplifies to
\onecolumngrid
\begin{eqnarray}
n_{1\uparrow/\downarrow}(t)&=&\frac{1}{g_{s}^{2}(4V_{12}^{2}+g_{s}^{2})} \left\{ n_{1\uparrow/\downarrow}(0)
\left[ \alpha_{1} \cos(\bar{s}_{1}t)- \alpha_{3} \cos(s_{3}t) \right]^{2}
+n_{2\uparrow/\downarrow}(0)V_{12}^{2} g_{s}^{2} \left[ \sin(\bar{s}_{1}t) + \sin(\bar{s}_{3}t) \right]^{2} \right.
\nonumber \\
&& + \left.
\left( 1-n_{1\downarrow/\uparrow}(0)\right) g_{s}^{2} \left[ \bar{s}_{1}\sin(\bar{s}_{1}t)-\bar{s}_{3}\sin(s_{3}t)\right]^{2}
+\left( 1-n_{2\downarrow/\uparrow}(0)\right) g_{s}^{2}V_{12}^{2} \left[\cos(\bar{s}_{1}t)-\cos(\bar{s}_{3}t)\right]^{2}\right\},
\label{eq:18} \\
%
n_{2\uparrow/\downarrow}(t)&=&\frac{1}{(4V_{12}^{2}+g_{s}^{2})} \left\{ n_{1\uparrow/\downarrow}(0) V_{12}^{2} \left[ \sin(\bar{s}_{1}t)+\sin(\bar{s}_{3}t)\right]^{2} + n_{2\uparrow/\downarrow}(0) \left[ \bar{s}_{3} \cos(\bar{s}_{1}t)+{\bar{s}_{1}}\cos(\bar{s}_{3}t) \right]^{2}
\right. \nonumber \\
&&\left. +\left( 1-n_{1\downarrow/\uparrow}(0)\right) V_{12}^{2} \left[ \cos(\bar{s}_{1}t)-\cos(\bar{s}_{3}t)\right]^{2}
+\left( 1-n_{2\downarrow/\uparrow}(0)\right) \left[ \bar{s}_{1}\sin(\bar{s}_{3}t)-\bar{s}_{3}\sin(\bar{s}_{1}t)\right]^{2}\right\},
\label{eq:19}
\end{eqnarray}
\twocolumngrid
\noindent
where $g_{s}=\frac{\Gamma_{S}}{2}$ and $\alpha_{1/3}=\frac{g_{s}^{2}}{2}\mp\frac{g_{s}}{2}\sqrt{g_{s}^{2}+4V_{12}^{2}}$. These expressions 
explicitly show an important role of the initial fillings.
One can notice that for some cases, e.g.\ when both quantum dots are initially singly occupied by the same spin, their occupancy is completely frozen, $n_{j\sigma}(t)=n_{j\sigma}(0)$. This is physically obvious, because electron occupying QD$_{1}$ is neither allowed to hop to QD$_{2}$ nor to the superconducting lead.

We now consider two different initial configurations, namely: (i) $n_{j\sigma}(0)=0$ or $n_{j\sigma}(0)=1$ and (ii) $n_{1\uparrow}=1$, $n_{1\downarrow}=0=n_{2\sigma}$. Note, that in the first case the electron transfer between QD$_{1}$ and superconducting lead is allowed right from the very beginning. Contrary to such scenario, in the second case any transfer of electron between the superconducting lead and QD$_{1}$ would be allowed only after  spin-$\uparrow$ electron jumps from QD$_{1}$ to QD$_{2}$. These initial conditions are effectively responsible for qualitatively different evolutions of  $n_{j\sigma}(t)$.

For the initially empty dots Eqs.\ (\ref{eq:18},\ref{eq:19}) imply
\begin{eqnarray}
n_{1/2\sigma}(t)&=&\frac{1}{4V_{12}^{2}+g_{s}^{2}} \left[ 4V_{12}^{2}\sin^{2}\left( (\bar{s}_{1}-\bar{s}_{3})t/2 \right)\right. \nonumber \\
&+&\alpha_{1/3}\sin^{2}(\bar{s}_{1}t)+\alpha_{3/1}\sin^{2}(\bar{s}_{3}t)],
\label{eq:20}
\end{eqnarray}
whereas for  $n_{j\sigma}(0)=1$ we obtain
\begin{eqnarray}
n_{1/2\sigma}(t)&=&\frac{1}{4V_{12}^{2}+g_{s}^{2}}[4V_{12}^{2}\cos^{2}\left( (\bar{s}_{1}-\bar{s}_{3})t/2\right)
\nonumber \\
&+&\alpha_{1/3}\cos^{2}(\bar{s}_{1}t)+\alpha_{3/1}\cos^{2}(\bar{s}_{3}t)] .
\label{eq:21}
\end{eqnarray}
We recognize here a superposition of three oscillations characterized by the periods $2\pi/|\bar{s}_{1}-\bar{s}_{3}|$, $\pi/|\bar{s}_{1}|$, and $\pi/|\bar{s}_{3}|$ with different amplitudes. In order to clarify such time-dependence  let us analyze the extreme cases when $V_{12}$ is much greater or smaller than $\Gamma_{S}$, respectively.

Expanding the contribution appearing in Eqs.\ (\ref{eq:20},\ref{eq:21}) in powers of $x\equiv\frac{\Gamma_{S}}{V_{12}}\ll 1$ up to the first non-vanishing terms, one obtains for the initially empty QDs
\begin{equation}
n_{1/2\sigma}(t)\simeq \sin^{2}\left( \frac{\Gamma_{S}}{4}t \right) \pm \frac{x}{8}  \sin\left( \frac{\Gamma_{S}}{2}t\right) \sin\left( \sqrt{4V_{12}^{2}+g_{s}^{2}}t\right)
\label{eq:22}
\end{equation}
and for the initially singly  occupied dots $n_{j\sigma}(0)=1$
\begin{equation}
n_{1/2\sigma}(t)\simeq  \cos^{2}\left( \frac{\Gamma_{S}}{4}t\right) \mp \frac{x}{8}  \sin\left( \frac{\Gamma_{S}}{2}t\right) \sin\left( \sqrt{4V_{12}^{2}+g_{s}^{2}}t\right) .
\label{eq:23}
\end{equation}
We notice, that $n_{j\sigma}(t)$ are governed mainly by the functions $\sin^{2}\left( \frac{\Gamma_{S}}{4}t \right)$ or $\cos^{2}\left( \frac{\Gamma_{S}}{4}t \right)$ with the period $T=4\pi/\Gamma_{S}$. One may argue, however, that for $V_{12}\gg\Gamma_{S}$ these occupancies should oscillate vs time in a way typical for a two-level system, characterized by the period $T=\pi/V_{12}$.
In fact such component is present here in the form of small correction, proportional to $\pm\frac{1}{8}\frac{\Gamma_{S}}{V_{12}}\sin(\frac{\Gamma_{S}}{2}t)$ with the period $\sim\frac{\pi}{V_{12}}$.
Some difference between the results obtained for the isolated two-level system in comparison to the present case manifests itself through influence of the initial occupancies of QDs.

Similar analysis for the opposite limit,  $\frac{1}{x} \ll 1$, yields
\begin{eqnarray}
n_{1/2\sigma}(t)&\simeq& \sin^{2}\left( \bar{s}_{3/1}t \right) + x^{-2} \left[ 16 \sin^{2}\left( \frac{\Gamma_{S}}{4}t\right) \right. \nonumber \\ &-& \left. 4 \sin^{2}\left( \bar{s}_{1/3}t\right) -12 \sin^{2}\left( \bar{s}_{3/1}t\right) \right],
\label{eq:xx}
\end{eqnarray}
for the initially empty dots, and
\begin{eqnarray}
n_{1/2\sigma}(t)&\simeq& \cos^{2}\left( \bar{s}_{3/1}t \right) + x^{-2} \left[ 16 \cos^{2}\left( \frac{\Gamma_{S}}{4}t\right) \right. \nonumber \\ &-& \left. 4 \cos^{2}\left( \bar{s}_{1/3}t\right) -12 \cos^{2}\left( \bar{s}_{3/1}t\right) \right],
\label{eq:yy}
\end{eqnarray}
for the initially filled dots. Time-dependent occupancy of QD$_{1}$ reveals the dominant quantum oscillations with period $T=2\pi/(\sqrt{4V_{12}^{2}+g_{s}^{2}}+g_{s})$. This result can be compared with the oscillations of a single quantum dot proximitized to superconducting lead, whose period is $2\pi/\Gamma_{S}$ \cite{Taranko-2018}. For the weak interdot coupling, the evolution of $n_{1\sigma}(t)$ is mainly affected by exchanging its electrons with the superconducting lead, whereas in the opposite case (for large $V_{12}$) both quantum dot occupancies can be partially exchanged. For this reason, $n_{2\sigma}(t)$ remarkably differs from $n_{1\sigma}(t)$ in the limit $\frac{V_{12}}{\Gamma_{S}}\ll 1$. The term $\sin^{2}\left( \bar{s}_{1}t \right)$ appearing in (\ref{eq:xx}) represents  oscillations with the period of
$2\pi /(\sqrt{4V_{12}^{2}+g^{2}_{S}}+g_{S})$ and the second term introduces corrections with the period of $4\pi/\Gamma_{S}$.

The aforementioned initial configuration with only single electron occupying QD$_{1}$ would imply quite different evolution of the considered system in comparison to the initially empty or filled both quantum dots. To illustrate it, we consider here the case when  at $t=0$ the single  electron, for instance $\uparrow$, occupies QD$_{1}$ (a neighbor of the superconducting lead). The time-dependent occupancies inferred from Eqs.~(\ref{eq:18},\ref{eq:19}) can be rewritten as follows
\begin{eqnarray}
n_{1\uparrow}(t)&=& \left[ 4V_{12}^{2}  \sin^{2}\left(\Gamma_{S}t/4 \right) + \Gamma_{S}^{2}/4 \right. \label{eq:xx27} \\
&+& \left. V_{12}^{2} \left( \cos\left( \bar{s}_{1}t\right) +  \cos\left( \bar{s}_{3}t\right) \right)^{2} \right]/a^{2} ,\nonumber \\
n_{1\downarrow}(t)&=& V_{12}^{2} \left( \cos\left(\bar{s}_{1}t \right)-\cos\left(\bar{s}_{3}t \right) \right)^{2}/a^{2} , \label{eq:xx28}\\
n_{2\uparrow}(t)&=& \left[ 4V_{12}^{2}  \sin^{2}\left(\Gamma_{S}t/4 \right) + \alpha_{1} \sin^{2}\left( \bar{s}_{3}t\right)  \right. \label{eq:xx29} \\
&+& \left. \alpha_{3} \sin^{2}\left( \bar{s}_{1}t\right) + V_{12}^{2} \left( \sin\left( \bar{s}_{1}t\right) +  \sin\left( \bar{s}_{3}t\right) \right)^{2} \right]/a^{2} , \nonumber \\
n_{2\downarrow}(t)&=& \left[  \alpha_{1} \sin^{2}\left( \bar{s}_{3}t\right) + \alpha_{3} \sin^{2}\left( \bar{s}_{1}t\right)  \right. \label{eq:xx30} \\
&+& \left.  V_{12}^{2} \left( \sin\left( \bar{s}_{1}t\right) -  \sin\left( \bar{s}_{3}t\right) \right)^{2} \right]/a^{2} , \nonumber
\end{eqnarray}
with $a=\sqrt{4V_{12}^{2}+g_{s}^{2}}$.

Let us consider the case of $V_{12}$ much larger than $\Gamma_{S}$. Performing similar calculations as those done for the initially both empty or filled QDs and ignoring the contributions proportional  and smaller than $(\Gamma_{S}/V_{12})^{2}$  one obtains
\begin{eqnarray}
n_{1\uparrow}(t)&\simeq & 1 - \cos^{2}\left(\Gamma_{S}t/4 \right) \sin^{2}\left( at/2 \right)  , \label{eq:xx31} \\
n_{1\downarrow}(t)&\simeq & \sin^{2}\left(\Gamma_{S}t/4 \right) \sin^{2}\left( at/2 \right) ,\label{eq:xx32}\\
n_{2\uparrow}(t)&\simeq & 1- \cos^{2}\left(\Gamma_{S}t/4 \right) \cos^{2}\left( at/2 \right) \label{eq:xx33} \\
&-&  \frac{x}{8} \sin\left( \Gamma_{S}t/2\right) \sin\left( at/2\right)  , \nonumber \\
n_{2\downarrow}(t)&\simeq & \sin^{2}\left(\Gamma_{S}t/4 \right) \cos^{2}\left( at/2 \right)  \label{eq:xx34} \\
&-& \frac{x}{8} \sin\left( \Gamma_{S}t/2\right) \sin\left( at/2\right)  . \nonumber
\end{eqnarray}
\begin{figure}
\centerline{\includegraphics[width=1\linewidth]{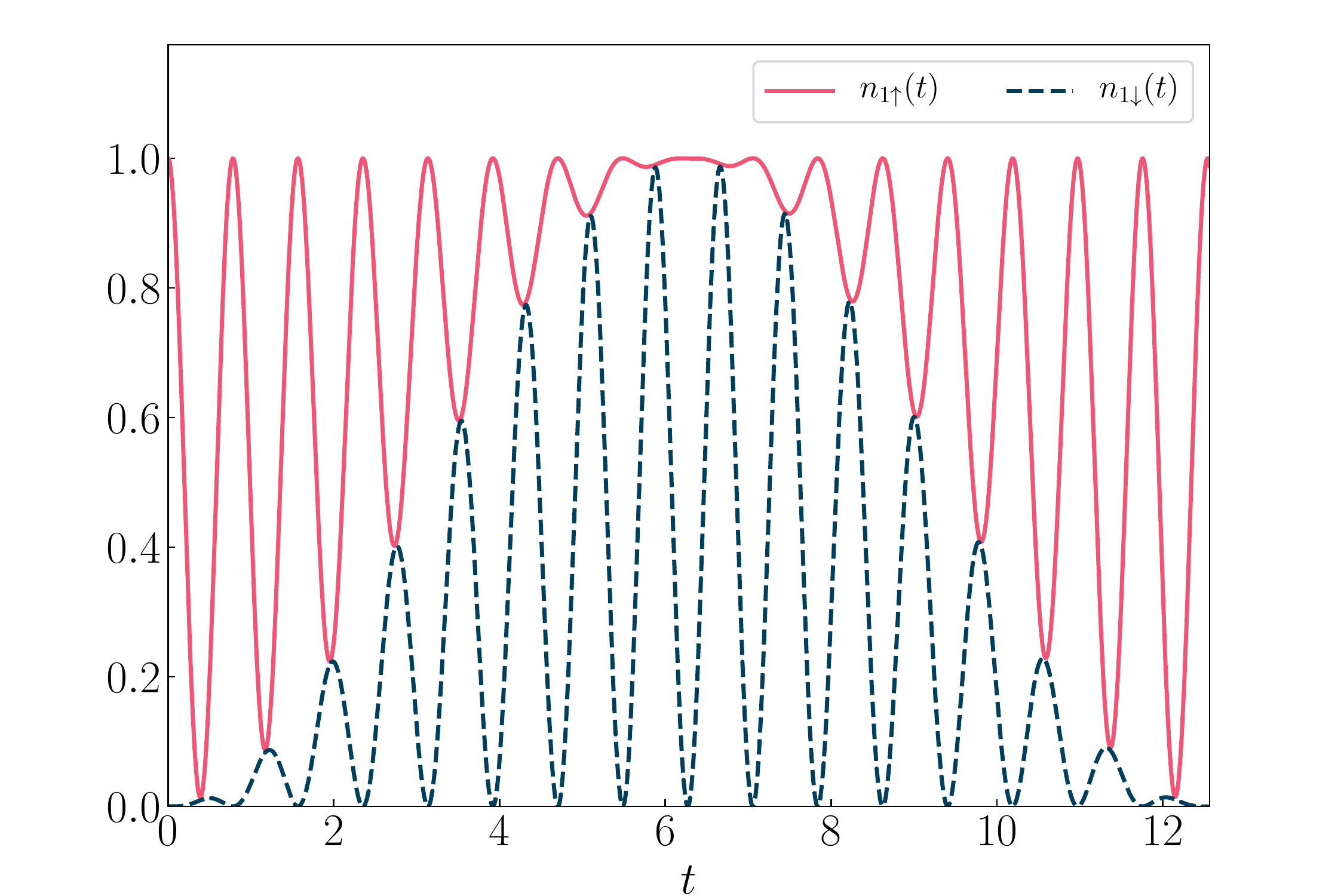}}
\caption{Time-dependent occupancies $n_{1\uparrow}(t)$/$n_{1\downarrow}(t)$ (shown by the solid/dashed lines) induced by the sudden switching of $\Gamma_{S}$ and interdot coupling $V_{12}$. Results are obtained for $\varepsilon_{j\sigma}=0$, $\Gamma_{N}=0$, $\Gamma_{S}=1$, $V_{12}=4$ assuming the initial configuration $n_{1\uparrow}(0)=1$ and $n_{1\downarrow}(0)=0=n_{2\sigma}(0)$. }
\label{fig2}
\end{figure}

Figure~\ref{fig2} shows  $n_{1\sigma}(t)$ obtained for $V_{12}=4\Gamma_{S}$, where we can clearly identify the quantum oscillations with the period equal to $ \pi/V_{12}$ typical for a two-level system. Their amplitude is modulated with other oscillations, whose period equal to $4\pi/\Gamma_{S}$ is controlled by the functions $\sin^{2}\left(\Gamma_{S}t/4 \right)$ or $ \mbox{\rm cos}^{2}\left( \Gamma_{S}t/4\right)$. Let us remark that in the case of single QD coupled to superconductor the time-dependent occupancy oscillates with the period twice shorter. Due to electron tunneling between the quantum dots and through interface between QD$_{1}$ and superconducting lead this period of quantum oscillations is present for all types of the initial configurations. Evolution of the second dot occupancy is similar to $n_{1\sigma}(t)$, therefore we skip its presentation.

\begin{figure}[b]
\centerline{\includegraphics[width=1\linewidth]{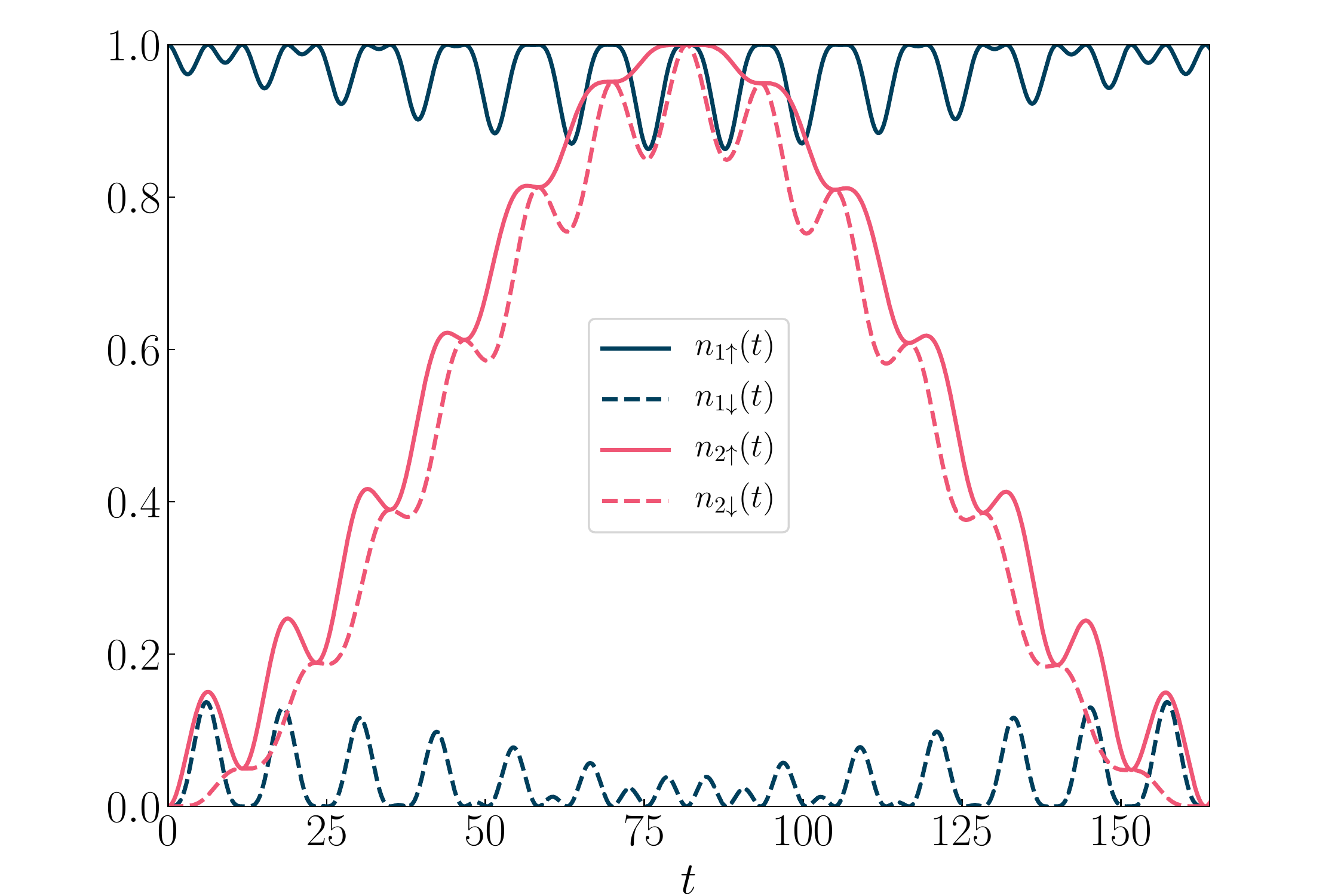}}
\caption{Time-dependent occupancies $n_{j\sigma}(t)$ [see the legend] obtained for the weak interdot coupling $V_{12}=0.1\Gamma_{S}$ using the same set of model parameters as in Fig.~\ref{fig2}.}
\label{fig3}
\end{figure}

We now expand the amplitudes of oscillating terms appearing in (\ref{eq:xx27}-\ref{eq:xx30}) to the second order in powers of $\frac{1}{x}$
\begin{eqnarray}
n_{1\uparrow}(t)&\simeq & 1 - 16 x^{-2} \cos^{2}\left(\Gamma_{S}t/4 \right) \sin^{2}\left( at/2 \right)  , \label{eq:xx35} \\
n_{1\downarrow}(t)&\simeq & 16 x^{-2} \sin^{2}\left(\Gamma_{S}t/4 \right) \sin^{2}\left( at/2 \right) ,\label{eq:xx36}\\
n_{2\uparrow}(t)&\simeq & \sin^{2}\left( \bar{s}_{1}t\right) + 8y^{2} \left[ 2 \sin^{2}\left(\Gamma_{S}t/4 \right) \right. \label{eq:xx37} \\
&-& \left. \sin^{2}\left( \bar{s}_{1}t\right) + \sin\left( \bar{s}_{1}t\right)\sin\left( \bar{s}_{3}t\right) \right]   , \nonumber \\
n_{2\downarrow}(t)&\simeq & \sin^{2}\left( \bar{s}_{1}t\right)  -  8x^{-2} \left[
\sin^{2}\left( \bar{s}_{1}t\right) \right.
\label{eq:xx38} \\
 &+& \left. \sin\left( \bar{s}_{1}t\right)\sin\left( \bar{s}_{3}t\right) \right]  . \nonumber
\end{eqnarray}
Note, that if one neglects all terms proportional to $x^{-2}$ then $n_{1\sigma}(t)$ would not change in time at all, irrespective  of its coupling to the second QD. QD$_{1}$ is initially  singly occupied by spin-$\uparrow$ electron which, at later time, might be transferred to QD$_{2}$. Such emptying  would enable  one of the Cooper pairs to leak from the superconducting reservoir onto QD$_{1}$ and, in next step, spin-$\downarrow$ electron could eventually be transferred onto QD$_{2}$. This reasoning explains why $n_{2\downarrow}(t)$ is slowly increasing right after the quench, owing to the terms proportional to $x^{-2}$ in Eq.~(\ref{eq:xx38}).

Figure~\ref{fig3} presents $n_{j\sigma}(t)$ obtained for the weak interdot coupling $V_{12}=0.1 \Gamma_{S}$. Differences between the occupancies of QD$_{1}$ and QD$_{2}$ are quite evident. QD$_{1}$ is nearly completely occupied/empty by $\uparrow$/$\downarrow$  electrons and such occupancy exhibits oscillations with the period $4\pi/\Gamma_{S}$ and small amplitude oscillating with another (larger) period  $2\pi/(\sqrt{4V_{12}^{2}+g_{s}^{2}}-g_{s})$. Time-dependent $n_{2\sigma}(t)$ is different, because the main contribution in Eqs.~(\ref{eq:xx37},\ref{eq:xx38}) simply oscillates with the period equal to $\pi/\bar{s}_{1}=2\pi/(\sqrt{4V^{2}_{12}+g^{2}_{S}}-g_{S})$ and its amplitude is 1. Further corrections, proportional to $x^{-2}$, introduce small variations of this amplitude, with the period $4\pi/\Gamma_{S}$.

\begin{figure}[h]
\centerline{\includegraphics[width=1\linewidth]{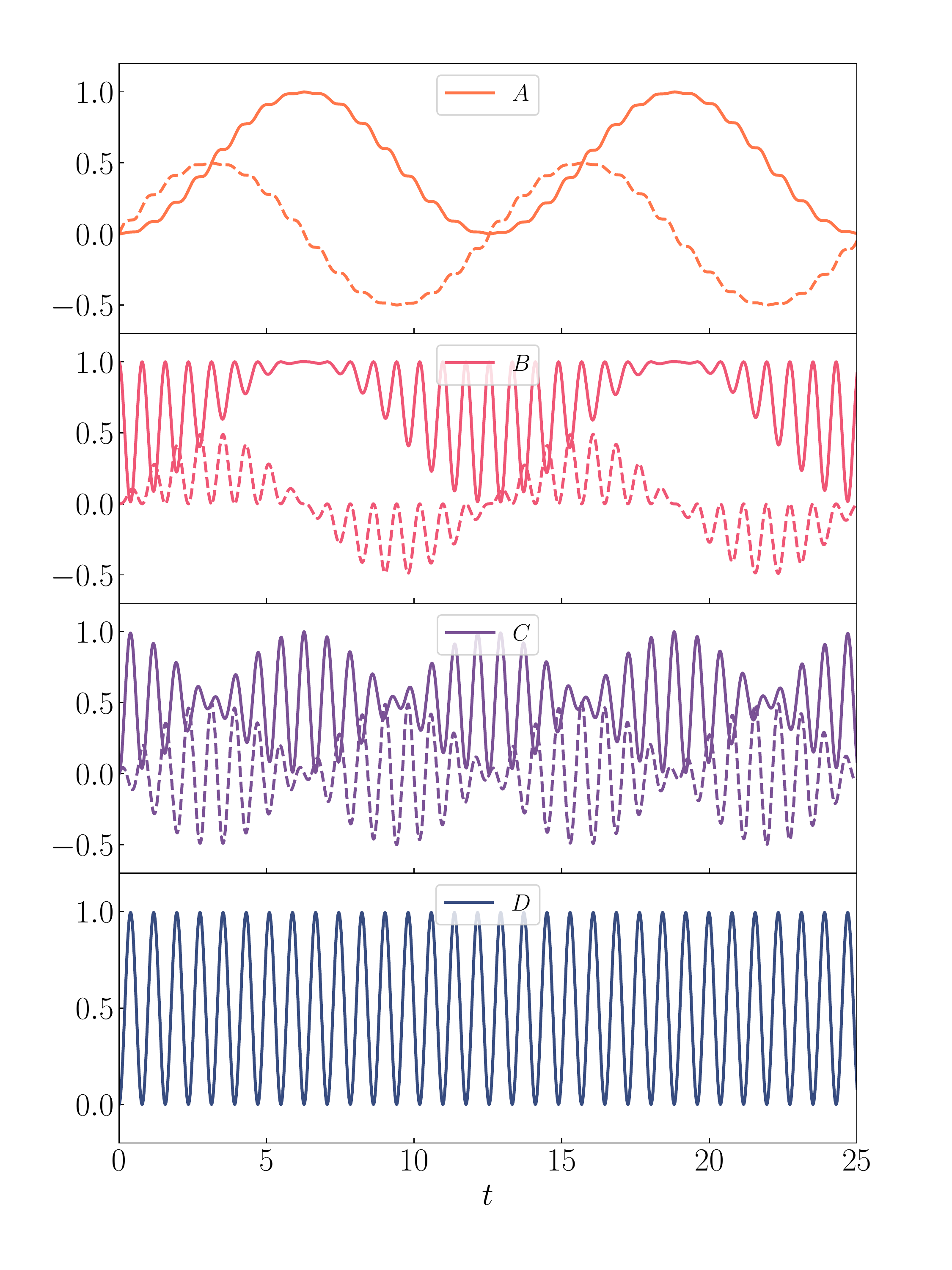}}
\caption{Time-dependent occupancy $n_{1\uparrow}(t)$ (solid lines) and charge current $j_{S\sigma}(t)$ (dashed curves) obtained for the strong interdot coupling $V_{12}=4\Gamma_{S}$ and several initial configurations (QD$_{2}$,QD$_{1}$): A=($0,0$), B=($0,\uparrow$), C=($\uparrow\downarrow,0$), and D=($\uparrow$,$\downarrow$), assuming $\Gamma_{S}=1$, $\Gamma_{N}=0$ and $\varepsilon_{j\sigma}=0$.}
\label{fig4}
\end{figure}

\begin{figure}[h]
\centerline{\includegraphics[width=1\linewidth]{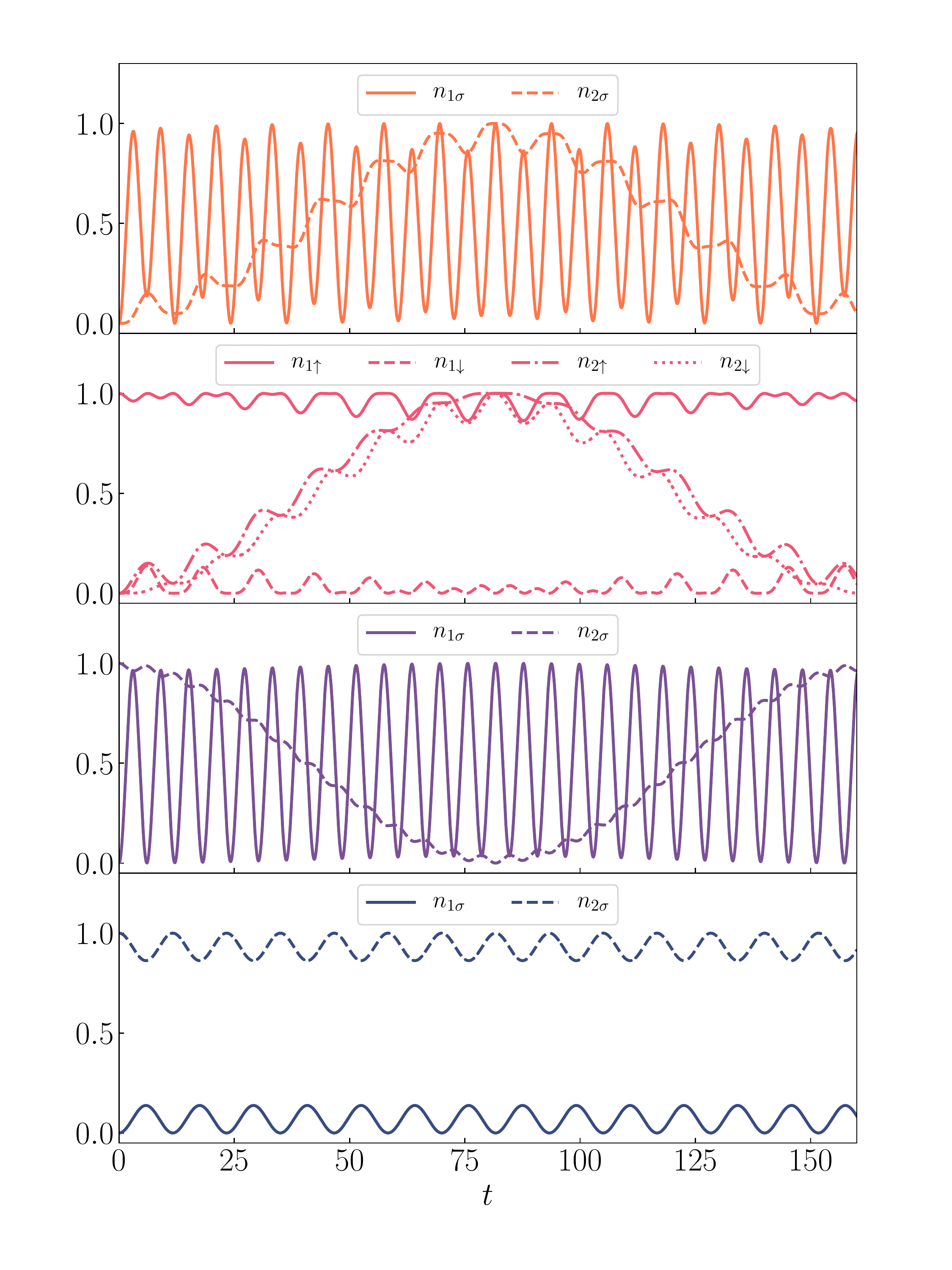}}
\caption{Time-dependent occupancy $n_{j\sigma}(t)$ (see the legend) obtained for the weak interdot coupling $V_{12}=0.1\Gamma_{S}$ and the same initial configuration as in Fig.~\ref{fig4}. Other parameters are: $\Gamma_{S}=1.0$, $\Gamma_{N}=0$ $\varepsilon_{j\sigma}=0$.}
\label{fig5}
\end{figure}

We have seen that the response of DQD to an abrupt coupling to the superconducting lead strongly depends on the initial fillings $n_{j\sigma}(0)$. It is sufficient to consider four representative types of the initial configurations in order to describe  all possible scenarios of the resulting $n_{j\sigma}(t)$ evolution. Figure\ \ref{fig4} shows $n_{1\uparrow}(t)$ obtained for the strong interdot coupling $V_{12}=4\Gamma_{S}$ for these initial conditions, namely (QD$_{2}$,QD$_{1}$)=($0,0$), ($0,\uparrow$), ($\uparrow\downarrow,0$), and ($\downarrow$,$\uparrow$). Oscillations with the period $\pi/V_{12}$ are well visible in all curves, whenever at $t= 0$ electrons occupy the QDs. Only for the case of the initially empty dots the oscillations have the period $4\pi/\Gamma_{S}$ with a small amplitude correction, exhibiting the period $\pi/V_{12}$. Specifically  for $n_{1\downarrow}(0)=1=n_{2\uparrow}(0)$, $n_{1\uparrow}(0)=0=n_{2\downarrow}(0)$ (Fig.\ref{fig4}D) we obtain
\begin{equation}
n_{1\uparrow}(t)= \frac{4V_{12}^{2}}{4V_{12}^{2}+g_{s}^{2}} \sin^{2}\left( \frac{\sqrt{4V_{12}^{2}+g_{s}^{2}}}{2} t \right) .
\label{eqx_38}
\end{equation}
For the large interdot coupling (\ref{eqx_38}) resembles the Rabi-type oscillations typical for a two-level quantum system.

In the opposite (small $V_{12}$) case for the initial D configuration we obtain
\begin{equation}
n_{1\uparrow}(t)= 16 \frac{V_{12}}{\Gamma_{S}} \sin^{2}\left( \frac{\Gamma_{S}}{4} t \right)
\end{equation}
with the period $4\pi/\Gamma_{S}$ (bottom panel in Fig.\ \ref{fig5}).
For the initial A and B configurations evolution considerably differs from the strong coupling limit.
Notice, that the time-dependent occupancies of both QDs are now completely different. The period  $2\pi/\Gamma_{S}$ shows up in $n_{1\sigma}(t)$ for the initial  A and C configurations. This period of oscillations can be assigned to the transfer of Cooper pairs back and forth from the superconducting lead onto QD$_{1}$. For the other (B and D) cases  we observe the oscillations with period $4\pi/\Gamma_{S}$  in the occupancies of both quantum dots.

For a deeper insight into the transient dynamics of the proximitized DQD we now consider the charge current $j_{S\sigma}(t)$ flowing from the superconducting lead to QD$_{1}$ and the inter-dot current $j_{12\sigma}(t)$, respectively. General expressions for $j_{S\sigma(t)}$ and $j_{12\sigma}(t)$ are presented in Appendix \ref{app:currents}. Here we focus on their values in the limit $\Gamma_{N}=0$
%
\onecolumngrid
\begin{eqnarray}
j_{S\sigma}(t)&=& \frac{1}{4V^{2}_{12}+g^{2}_{s}} \left\{ 2\left[ 1-n_{1\sigma}(0)-n_{1-\sigma}(0)  \right] \left[\bar{s}_{1}\sin(\bar{s}_{1}t)-\bar{s}_{3}\sin(\bar{s}_{3}t) \right] \left[\alpha_{1}\cos(\bar{s}_{1}t)-\alpha_{3}\cos(\bar{s}_{3}t) \right] \right. \nonumber \\
&+& \left. \Gamma_{S}V^{2}_{12}\left[ 1-n_{2\sigma}(0)-n_{2-\sigma}(0)\right]\left[\cos(\bar{s}_{1}t)-\cos(\bar{s}_{3}t)\right]\left[\sin(\bar{s}_{1}t)+\sin(\bar{s}_{3}t)\right]   \right\} ,
\label{eq:xx41}
\\
j_{12\uparrow/\downarrow}(t)&=& \frac{2V^{2}_{12}}{g_{s}(4V^{2}_{12}+g^{2}_{s})} \left\{ n_{1\uparrow/\downarrow}(0)\left[\alpha_{3}\cos(\bar{s}_{3}t)- \alpha_{1}\cos(\bar{s}_{1}t)\right]\left[\sin(\bar{s}_{1}t)+\sin(\bar{s}_{3}t)\right] \right. \nonumber \\
&-&n_{2\uparrow/\downarrow}(0)\left[ \alpha_{3}\cos(\bar{s}_{1}t)- \alpha_{1}\cos(\bar{s}_{3}t)\right]\left[\sin(\bar{s}_{1}t)+\sin(\bar{s}_{3}t)\right] \nonumber \\
&+&g_{s}\left[1-n_{1\downarrow/\uparrow}(0)\right]\left[\bar{s}_{1}\sin(\bar{s}_{1}t)-\bar{s}_{3}\sin(\bar{s}_{3}t) \right] \left[ \cos(\bar{s}_{3}t)- \cos(\bar{s}_{1}t)\right] \nonumber \\
&+& \left. g_{s}\left[1-n_{2\downarrow/\uparrow}(0)\right]\left[\bar{s}_{1}\sin(\bar{s}_{3}t)-\bar{s}_{3}\sin(\bar{s}_{1}t) \right] \left[ \cos(\bar{s}_{3}t)- \cos(\bar{s}_{1}t)\right] \right\}.
\label{eq:xx42}
\end{eqnarray}
\twocolumngrid

Broken lines in Fig.~\ref{fig4} display the currents $j_{S\sigma}(t)$  obtained for several initial conditions and the strong interdot coupling, $V_{12}=4\Gamma_{S}$. We observe that the time-dependence of the current $j_{S\sigma}(t)$ resembles the evolution of $n_{j\uparrow}(t)$, because they are linked through the charge conservation law. In particular, for the initial B and C configurations we recognize the oscillations with period $T\simeq \pi/V_{12}$, which are modulated by the envelope function oscillating with another period $T=4\pi/\Gamma_{S}$. Contrary to such behavior, for the initially empty dots the time-dependent current $j_{S\sigma}(t)$ is strictly governed by $\sin^{2}(\frac{\Gamma_s}{4}t)$ with the period $T=4\pi/\Gamma_{S}$.

Eqs.\ (\ref{eq:xx41},\ref{eq:xx42}) imply under what initial configuration (QD$_{2}$,QD$_{1}$) the charge current $j_{S\sigma}(t)$ can eventually vanish. For the case  $(\sigma,\sigma)$ the charge tunneling is neither allowed to flow from QD$_{1}$ to the neighboring QD$_{2}$ nor to the superconducting lead, so in consequence the occupancies of DQDs would be frozen. For the other configuration  $(\sigma,\bar{\sigma})$ this behavior would not be observed, because spin-$\downarrow$ (spin-$\uparrow$) electron can tunnel from the first to the second quantum dot simultaneously with the Cooper pair transmittance from the superconducting lead onto QD$_{1}$. In the latter case
the finite $j_{12\sigma}(t)$ and vanishing $j_{S\sigma}(t)$ currents could be observed.

The initial $( \uparrow,\downarrow )$ or $( \downarrow,\uparrow )$ configurations evolve in time through the intermediate states $( \uparrow\downarrow,0 )$, $( 0,\uparrow\downarrow )$, $( \uparrow\downarrow,\uparrow\downarrow )$, $( 0,0 )$, $( \uparrow,\downarrow )$ and $( \downarrow,\uparrow )$, respectively. It can  be shown, by solving the time-dependent Schr\"{o}dinger equation, that at arbitrary time the double quantum dot can be found  with equal probabilities in the configurations $(0,\uparrow\downarrow)$, $(\uparrow\downarrow,0)$ or with equal probabilities in the configurations $(0,0)$, $(\uparrow\downarrow,\uparrow\downarrow)$. It means that in both cases  the electron pairs can tunnel with the same probability from QD$_{1}$ either to the superconducting lead or in the opposite direction. In consequence, the current $j_{S\sigma}(t)$ vanishes.
This conclusion can be also formally inferred from Eq.~(\ref{eq:xx41}).

For the weak interdot coupling $V_{12}$  and assuming the initial conditions  ($0,0$) or ($\uparrow\downarrow,0)$, the current $j_{S\sigma}(t)$ evolves with respect to time in a way similar to the occupancy $n_{1\sigma}(t)$ being characterized by the quantum oscillations with period $2\pi/\Gamma_{S}$ and approximately constant amplitude. For the initial conditions ($0,\uparrow$) the current $j_{S\sigma}(t)$ oscillates with the period $4\pi/\Gamma_{S}$, in analogy to time-dependent $n_{1\sigma}(t)$ displayed in Figs. \ref{fig3} and \ref{fig5}.

We now briefly consider the on-dot $\langle \hat{c}_{j\downarrow}(t)\hat{c}_{j\uparrow}(t) \rangle$ and inter-dot $\langle \hat{c}_{1\downarrow}(t)\hat{c}_{2\uparrow}(t) \rangle$  pairings, whose general expressions are presented in Appendix \ref{app:currents}.
In the limit of $\Gamma_{N}=0$ their simplified analytical expression are given by
%
\onecolumngrid
\begin{eqnarray}
\langle \hat{c}_{1\downarrow}(t)\hat{c}_{1\uparrow}(t) \rangle &=& \frac{i}{g_{s}(4V^{2}_{12}+g^{2}_{s})} \left\{ \left[ 1-n_{1\uparrow}(0)-n_{1\downarrow}(0)\right] \left[ \bar{s}_{3}\sin(\bar{s}_{3}t)-\bar{s}_{1}\sin(\bar{s}_{1}t) \right] \left[ \alpha_{1}\cos(\bar{s}_{1}t)-\alpha_{3}cos(\bar{s}_{3}t)\right] \right. \nonumber \\
&+& \left. g_{s}V^{2}_{12}\left[ 1-n_{2\uparrow}(0)-n_{2\downarrow}(0) \right] \left[ \cos(\bar{s}_{3}t)-\cos(\bar{s}_{1}t)\right]\left[ \sin(\bar{s}_{1}t)+\sin(\bar{s}_{3}t) \right] \right\} ,
\label{eq:xx43} \\
\langle \hat{c}_{2\downarrow}(t)\hat{c}_{2\uparrow}(t) \rangle &=& \frac{iV^{2}_{12}}{4V^{2}_{12}+g^{2}_{s}} \left\{ \left[ 1-n_{1\uparrow}(0)-n_{1\downarrow}(0) \right] \left[ \cos(\bar{s}_{1}t)-\cos(\bar{s}_{3}t) \right] \left[ \sin(\bar{s}_{1}t)+\sin(\bar{s}_{3}t) \right] \right. \nonumber \\
&+&\left. \left[ 1-n_{2\uparrow}(0)-n_{2\downarrow}(0) \right] \left[ \bar{s}_{1}\sin(\bar{s}_{3}t)-\bar{s}_{3}\sin(\bar{s}_{1}t) \right] \left[ \bar{s}_{3}\cos(\bar{s}_{1}t)+\bar{s}_{1}\cos(\bar{s}_{3}t) \right] \right\}/V^{2}_{12} ,
\label{eq:xx44} \\
\langle \hat{c}_{1\downarrow}(t)\hat{c}_{2\uparrow}(t) \rangle &=& \frac{V_{12}}{4V^{2}_{12}+g^{2}_{s}} \left\{ n_{1\uparrow}(0)\left[ \bar{s}_{3}\sin(\bar{s}_{3}t)-\bar{s}_{1}\sin(\bar{s}_{1}t)\right] \left[ \sin(\bar{s}_{1}t)+\sin(\bar{s}_{3}t) \right] \right. \nonumber \\
&+&\frac{1}{g_{s}}n_{2\uparrow}(0)\left[ \cos(s_{1}t)-\cos(s_{3}t) \right] \left[ \alpha_{1}\cos(\bar{s}_{3}t)-\alpha_{3}\cos(\bar{s}_{1}t)\right] \nonumber \\
&+&\frac{1}{g_{s}}\left[ 1-n_{1\downarrow}(0) \right] \left[ \cos(s_{1}t)-\cos(s_{3}t)\right] \left[ \alpha_{1}\cos(\bar{s}_{1}t)-\alpha_{3}\cos(\bar{s}_{3}t)\right] \nonumber \\
&+& \left. \left[ 1-n_{2\downarrow}(0)\right] \left[ \bar{s}_{1}\sin(\bar{s}_{3}t)-\bar{s}_{3}\sin(\bar{s}_{1}t)\right] \left[\sin(\bar{s}_{1}t)+\sin(\bar{s}_{3}t)\right] \right\} .
\label{eq:xx45}
\end{eqnarray}
\twocolumngrid
We notice that the on-dot pairing functions (\ref{eq:xx43},\ref{eq:xx44}) are purely imaginary whereas the inter-dot pairing function (\ref{eq:xx45}) is real. They eventually vanish when each QD is initially singly occupied by the same spin electron. Let us recall, that under such circumstances the current $j_{S\sigma}(t)$ vanishes as well. In contrast to this situation, when QDs are initially singly occupied by electrons of opposite spins,
then the on-dot pairing functions (\ref{eq:xx43},\ref{eq:xx44}) vanish, whereas the inter-dot pairing  (\ref{eq:xx45}) survives. We have checked
that for arbitrary situations the following relationship $j_{S\sigma}(t)=-\Gamma_{S} \mbox{\rm Im} \langle \hat{c}_{1\downarrow}(t)\hat{c}_{1\uparrow}(t) \rangle$ is obeyed. This identity
has been widely used in studies of the Josephson current under the stationary conditions \cite{Zonda-2015,Zonda-2016}.

\bibliography{biblio_2dots}

\end{document}